\documentclass[prl,showpacs,twocolumn,superscriptaddress,floatfix,10pt]{revtex4-1}
\usepackage{graphicx}
\usepackage{amsfonts,amssymb,stmaryrd,latexsym,amsmath}
\usepackage[colorlinks,citecolor=blue,urlcolor=blue,linkcolor=blue]{hyperref}
\usepackage{slashed}
\usepackage{orcidlink}
\usepackage{bm}
\usepackage{multirow}
\usepackage{bbm}
\usepackage[normalem]{ulem}

\allowdisplaybreaks

\renewcommand{\arraystretch}{1.8}

\begin{document}



\title{Uncovering the mystery of $X(3872)$ with the coupled-channel dynamics}
\author{Jun-Zhang Wang\,\orcidlink{0000-0002-3404-8569}}\email{wangjzh2022@pku.edu.cn}
\affiliation{School of Physics and Center of High Energy Physics, Peking University, Beijing 100871, China}
\affiliation{Department of Physics, Chongqing University, Chongqing 401331, China}

\author{Zi-Yang Lin\,\orcidlink{0000-0001-7887-9391}}\email{lzy$_$15@pku.edu.cn}
\affiliation{School of Physics and Center of High Energy Physics, Peking University, Beijing 100871, China}

\author{Yan-Ke Chen\,\orcidlink{
0000-0002-9984-163X}}\email{chenyanke@stu.pku.edu.cn}
\affiliation{School of Physics and Center of High Energy Physics, Peking University, Beijing 100871, China}

\author{Lu Meng\,\orcidlink{0000-0001-9791-7138}}\email{lu.meng@rub.de}
\affiliation{Institut f\"ur Theoretische Physik II, Ruhr-Universit\"at Bochum,  D-44780 Bochum, Germany }

\author{Shi-Lin Zhu\,\orcidlink{0000-0002-4055-6906}}\email{zhusl@pku.edu.cn
}
\affiliation{School of Physics and Center of High Energy Physics, Peking University, Beijing 100871, China}

\date{\today}

\begin{abstract}
The $X(3872)$, as the first and the most crucial member in the exotic charmoniumlike $XYZ$ family, has been studied for a long time. However, its dynamical origin, whether stemming from a $D\bar{D}^*$ hadronic molecule or the first excited $P$-wave charmonium $\chi_{c1}(2P)$, remains controversial. In this Letter, we demonstrate that the $X(3872)$ definitely does not result from the mass shift of {the  higher bare $\chi_{c1}(2P)$ resonance pole } in the coupled-channel dynamics involving a short-distance $c\bar{c}$ core and the long-distance $D\bar{D}^*$ channels. { Instead, it originates from either the $D\bar{D}^*$ molecular pole or the shadow pole associated with the anti-resonance of the $P$-wave charmonium, depending on the weak or strong coupling mode, respectively. To differentiate these origins and fully exploit the nature of $X(3872)$, we conduct a comprehensive analysis in a couple-channel dynamics framework, including the isospin violation, the three-body $D\bar{D}\pi$ effect, the dynamical width of $D^*$, and non-open-charm decays of the bare $\chi_{c1}(2P)$. Our findings highlight the pivotal role of the coupled-channel dynamics in explaining the disparity between the pole widths of $X(3872)$ and $T_{cc}^+$, while also predicting a new resonance with $J^{PC}=1^{++}$ around 4.0 GeV. By matching the newly observed $\chi_{c1}(4010)$ by the LHCb Collaboration  to our predicted resonance, we conclude that the $X(3872)$ most likely originates from the $D\bar{D}^*$ pole with a confidence level exceeding $99.7\%$. } 
\end{abstract}

\maketitle

{\it Introduction.}--- { Over the past two decades, one of the central challenges in the strong interaction physics is to understand numerous exotic $XYZ$ resonances 
(see Refs. \cite{Chen:2016qju,Liu:2019zoy,Guo:2017jvc,Brambilla:2019esw,Olsen:2017bmm,Meng:2022ozq} for recent progresses). As the undisputed superstar among them, the nature of $X(3872)$ is the most mysterious, which is mainly manifested in that its mass exactly coincides to the threshold of $D^{0}\bar{D}^{*0}$/$D^{*0}\bar{D}^{0}$ and its decay pattern exhibits a strong coupling to the $D^0\bar{D}^0\pi^0$ final state \cite{Belle:2003nnu,Belle:2006olv}. These distinctive features imply the central role of the $D\bar{D}^*$ (specifically refers to $D\bar{D}^*/\bar{D}D^*$) interaction in the generation of $X(3872)$, leading to the natural proposal of a hadronic molecular interpretation of $X(3872)$ \cite{Voloshin:2003nt,Braaten:2003he,Swanson:2003tb,Tornqvist:2004qy,Fleming:2007rp,Liu:2008fh}. { In a related development, the LHCb Collaboration recently discovered the doubly charmed tetraquark $T_{cc}^+$ in the prompt production
of the $pp$ collision \cite{LHCb:2021vvq,LHCb:2021auc}, which has been widely recognized as an ideal candidate for a shallow $DD^*$ bound state }\cite{Manohar:1992nd,Janc:2004qn,Ohkoda:2012hv,Li:2012ss,Chen:2021cfl,Dong:2021bvy,Feijoo:2021ppq,Albaladejo:2021vln,Fleming:2021wmk,Meng:2021jnw,Du:2021zzh,Lin:2022wmj,Cheng:2022qcm,Ke:2021rxd,Ling:2021bir,Liu:2019yye,Yan:2021wdl,Jin:2021cxj,Xin:2021wcr,Shi:2022slq,Ortega:2022efc,Du:2023hlu,Wang:2023iaz,Wang:2022jop,Chen:2023fgl,Chen:2021vhg}. }

In contrast to the doubly charmed tetraquark $T_{cc}^+$, the nature of $X(3872)$  presents a more intricate challenge, as its $D\bar{D}^*$ component inevitably interacts with the conventional charmonium states with the same quantum number $J^{PC}=1^{++}$ \cite{LHCb:2013kgk,LHCb:2015jfc}, most notably the nearby $\chi_{c1}(2P)$ state (hereinafter denoted as $\chi_{c1}^{\prime}$). This dynamics indicates that the realistic wave function of $X(3872)$ exhibits dual characteristics: a compact charmonium in the short-distance region and a loose $D\bar{D}^*$ structure in the long-distance region \cite{Baru:2010ww,Hanhart:2011jz,Hammer:2016prh,Hanhart:2022qxq}. In light of this, a widely held opinion suggests that the $X(3872)$ arises from the mass shift of the higher bare $\chi_{c1}^{\prime}$ state, driven by the unquenched $D\bar{D}^*$ loop correction to the propagator of the $P$-wave charmonium \cite{Pennington:2007xr,Ono:1983rd,Li:2009ad,Zhou:2013ada,Duan:2020tsx,Deng:2023mza,Man:2024mvl}. Consequently, whether the $X(3872)$ is a hadronic molecule or an excited charmonium remains controversial. Regarding this long-standing view, in this Letter, we demonstrate that the $X(3872)$ definitely does not originate from the bare $\chi_{c1}^{\prime}$ resonance pole in the coupled-channel dynamics involving the $c\bar{c}$ core and the $D\bar{D}^*$ channels. 
{ In contrast}, the $X(3872)$ stems from either the $D\bar{D}^*$ pole in the so-called weak-coupling mode or the shadow pole associated with anti-resonance of charmonium in the strong-coupling mode. { Therefore, our research has rectified the misunderstanding regarding the pole origin  of $X(3872)$ in previous literature \cite{Pennington:2007xr,Ono:1983rd,Li:2009ad,Zhou:2013ada,Duan:2020tsx,Deng:2023mza,Man:2024mvl} and concluded that the conventional charmonium explanation of $X(3872)$ can be ruled out. }   

{ To fully understand the nature of $X(3872)$, we perform a coupled-channel dynamical calculation, incorporating the $D\bar{D}^*$ interaction via the chiral effective field theory (ChEFT) and three key mechanisms contributing to the pole width of $X(3872)$: the $D\bar{D}\pi$ three-body cut from the one-pion exchange (OPE) in $D\bar{D}^*$ scattering,  the dynamical width of $\bar{D}^*$ and the non-open-charm decays of the bare $\chi_{c1}^{\prime}$ state. A notable prediction is the existence of a new resonance with $J^{PC}=1^{++}$ around 4.0 GeV when the $X(3872)$ pole emerges. Remarkably, we establish a quantitative connection among the dynamical origin of $X(3872)$, its pole width and the properties of this predicted resonance, which offers a novel approach to distinguish the origins of $X(3872)$. Thanks to the LHCb observation of $\chi_{c1}(4010)$ { and given our assumptions of $\chi_{c1}(4010)$ as the predicted new resonance}, we conclude with over $99.7\%$ confidence that $X(3872)$ originates from the $D\bar{D}^*$ pole. Our findings
demonstrate the irreplaceable role of coupled-channel dynamics in unlocking the mystery of $X(3872)$, while providing important implications for understanding other $XYZ$ states. }

{\it Pole evolution of $X(3872)$ in the coupled-channel dynamics.}---We adopt a coupled-channel framework to study the properties of $X(3872)$, in which the $D\bar{D}^*$ scattering dynamics is also included.
The hadronic fock state can be written as
\begin{align}
| \Phi \rangle=c_0 | \Phi_0 \rangle+\sum_i\int \frac{d^3\textbf{q}}{(2\pi)^3} \phi_i(\textbf{q})| \Phi_i \rangle_{\textbf{q}},
\end{align}
where $| \Phi_0 \rangle$ and $| \Phi_i \rangle_{\textbf{q}}$ correspond to the bare fock state and the $i$-channel hadron-hadron continuum associated with the relative momentum $\textbf{q}$, respectively. { Due to poor knowledge of the compact tetraquark $[c\bar{c}q\bar{q}]$ spectrum in the present theoretical investigations, it seems rather unlikely that we will explore the role of this bare fock state in the coupled channel dynamics associated with the formation of $X(3872)$. In order to clearly identify this point, more experimental hints and evidence from Lattice QCD are needed. } Then the coupled-channel Schr\"{o}dinger equation can be written as 
\begin{align}
\begin{pmatrix}
    \mathcal{H}_0 & \mathcal{H}_{01} & \mathcal{H}_{02} & \dots \\
    \mathcal{H}_{10} & \mathcal{H}_1 & \mathcal{H}_{12} & \dots\\
    \mathcal{H}_{20} & \mathcal{H}_{21} & \mathcal{H}_{2} & \dots\\
   \vdots  &  \vdots &  \vdots & \ddots\\
\end{pmatrix}
\begin{pmatrix}
    c_0 | \Phi_0 \rangle  \\
  | \Phi_1 \rangle  \\
| \Phi_2 \rangle   \\
\vdots \\
\end{pmatrix}
&=E \begin{pmatrix}
    c_0 | \Phi_0 \rangle  \\
  | \Phi_1 \rangle  \\
| \Phi_2 \rangle   \\
\vdots \\
\end{pmatrix}
\end{align}
with 
\begin{align}
    | \Phi_i \rangle=\int \frac{d^3\textbf{q}}{(2\pi)^3} \phi_i(\textbf{q})| \Phi_i \rangle_{\textbf{q}}, \quad(i=1,2,\cdots).
\end{align}
After expanding the above matrix equation and applying the relation $\langle \Phi_0| \mathcal{H}_0 | \Phi_0 \rangle=M_0 $, $\langle \Phi_j |_{\textbf{q}} \mathcal{H}_{j0} | \Phi_0 \rangle=\mathcal{V}_{0j}(\textbf{q})$ and $\langle \Phi_j |_{\textbf{q}^{\prime}} \mathcal{H}_{ji} | \Phi_i  \rangle _{\textbf{q}}=\delta_{ij}E_k + V_{ij}(\textbf{q},\textbf{q}^{\prime})$, one can obtain
\begin{align}
&\, \sum_i \int \phi_i(\textbf{q}) \mathcal{V}_{i0}(\textbf{q})  \frac{d^3\textbf{q}}{(2\pi)^3}=(E-M_0)c_0,  \label{eq6}\\ 
 &\, c_0\mathcal{V}_{0j}(\textbf{q})+\sum_i\int(\delta_{ij}E_k + V_{ij}(\textbf{q},\textbf{q}^{\prime}))\phi_i(\textbf{q}^{\prime}) \frac{d^3\textbf{q}^{\prime}}{(2\pi)^3} \nonumber\\
 &\,\hspace{0.3\hsize} =E \phi_j(\textbf{q})~~~(j=1,2,3 \dots),  \label{eq7}
\end{align}
respectively, where $M_0$ is the bare mass of $| \Phi_0 \rangle$, the $\mathcal{V}_{0j}$ corresponds to the transition amplitude between the bare state and the $j$-th channel, and the $E_k$ and $V_{ij}$ stand for the kinetic term $\textbf{q}^2/(2\mu)$ and the potential of the corresponding hadron-hadron scattering, respectively. Combining  Eqs.~(\ref{eq6}) and (\ref{eq7}), we have
\begin{align}
 &\,  \sum_i(\delta_{ij}E_k +\int( V_{ij}(\textbf{q},\textbf{q}^{\prime})+ \mathcal{V}_{ij}(\textbf{q},\textbf{q}^{\prime}))\phi_i(\textbf{q}^{\prime}) \frac{d^3\textbf{q}^{\prime}}{(2\pi)^3}) \nonumber\\
 &\,\hspace{0.3\hsize} =E\phi_j(\textbf{q})~~~(j=1,2,3 \dots), \label{eq8}
\end{align}
where 
$\mathcal{V}_{ij}(\textbf{q},\textbf{q}^{\prime})$=$\frac{\mathcal{V}_{i0}(\textbf{q}^{\prime})\mathcal{V}_{0j}(\textbf{q})}{E-M_0+i\epsilon}$.
Equation (\ref{eq8}) is a standard coupled-channel Schr\"{o}dinger equation in momentum space. The coupled-channel problem between the bare state and the hadronic continuum is converted to a hadron-hadron scattering problem including an extra $s$-channel effective potential  $\mathcal{V}_{ij}(\textbf{q},\textbf{q}^{\prime})$.

We consider a two-channel calculation involving the $[D^0\bar{D}^{*0}]$ and $[D^+\bar{D}^{*-}]$, where the square brackets are the shorthands of the $C=+$ states such as $[D^0\bar{D}^{*0}]$=$\frac{1}{\sqrt{2}}(D^0\bar{D}^{*0}-\bar{D}^0D^{*0})$. Because the mass of $X(3872) $ lies close to the threshold of $D^0\bar{D}^{*0}$, the $D\bar{D}^* \to D\bar{D}^*$ scattering should be governed by the leading-order  contact and OPE interactions  in the effective field theory \cite{Lin:2022wmj}.  The total effective potential can be written as
\begin{align} 
V_{\text{Total}}(\textbf{q},\textbf{q}^{\prime})=
\begin{pmatrix}
    C_t-V_{\pi^0}+\mathcal{V}_{11} & C_t^{\prime}-2V_{\pi^{\pm}}+\mathcal{V}_{12} \\
  C_t^{\prime}-2V_{\pi^{\pm}}+\mathcal{V}_{21} & C_t-V_{\pi^{0}}+\mathcal{V}_{22}\\
 \end{pmatrix}, \label{eq9}
\end{align} 
where $C_t$ and $C_t^{\prime}$  are the contact terms, $V_{\pi}$ is the OPE potential and $\mathcal{V}_{11}=\mathcal{V}_{12}=\mathcal{V}_{21}=\mathcal{V}_{22}$. Since the ChEFT only works at small momentum, we introduce a Gaussian cutoff to regularize the contact and OPE potential, i.e., 
\begin{align}
\mathcal{F}(\textbf{q},\textbf{q}^{\prime})=\text{exp}(-(\textbf{q}^2+\textbf{q}^{\prime 2})/\Lambda^2),
\end{align}
where the cutoff $\Lambda=0.5$ GeV is usually taken for the $DD^*$ or $D\bar{D}^*$ system \cite{Lin:2022wmj}.
In order to explore the trajectory of the pole generated in  the $D\bar{D}^* \to D\bar{D}^*$ scattering, we first ignore the OPE contribution and adopt the momentum-dependent form of 
\begin{align}
\mathcal{V}=\frac{g^2}{2M_0} e^{-(\textbf{q}^2+\textbf{q}^{\prime 2})/\alpha^2}/(E-M_0+i\epsilon),   
\end{align}
where $\alpha$ and $g$ characterize the coupling range and strength between $\chi_{c1}^{\prime}$ and $D\bar{D}^{*}$, respectively. Then the effective potential in Eq.~(\ref{eq9}) becomes separable, and it would be very convenient to analytically solve the Lippmann-Schwinger equation (LSE) and search for the poles of the $T$ matrix \cite{Oller:1998hw}.

The general coupled-channel LSE is given by
\begin{align}
  T^{\alpha}_{\beta}(q,q^{\prime})\bm{=}V^{\alpha}_{\beta}(q,q^{\prime})+\sum_{\gamma}\int_0^{\infty}\frac{dk k^2}{(2\pi)^3}\frac{V^{\alpha}_{\gamma}(q,k)T^{\gamma}_{\beta}(k,q^{\prime})}{E-k^2/(2\mu_{\gamma})},    \label{eq11} 
 \end{align}
where $V^{\alpha}_{\beta}(q,q^{\prime})$ is the partial-wave-projected potential from the $\alpha$ channel to the $\beta$ channel, and $q$ and $q^{\prime}$ correspond to the initial and final momentum, respectively. The reduced mass $\mu_{\gamma}\bm{=}m_1m_2/(m_1+m_2)$.

\begin{figure}
    \centering
    \includegraphics[width=0.46\textwidth]{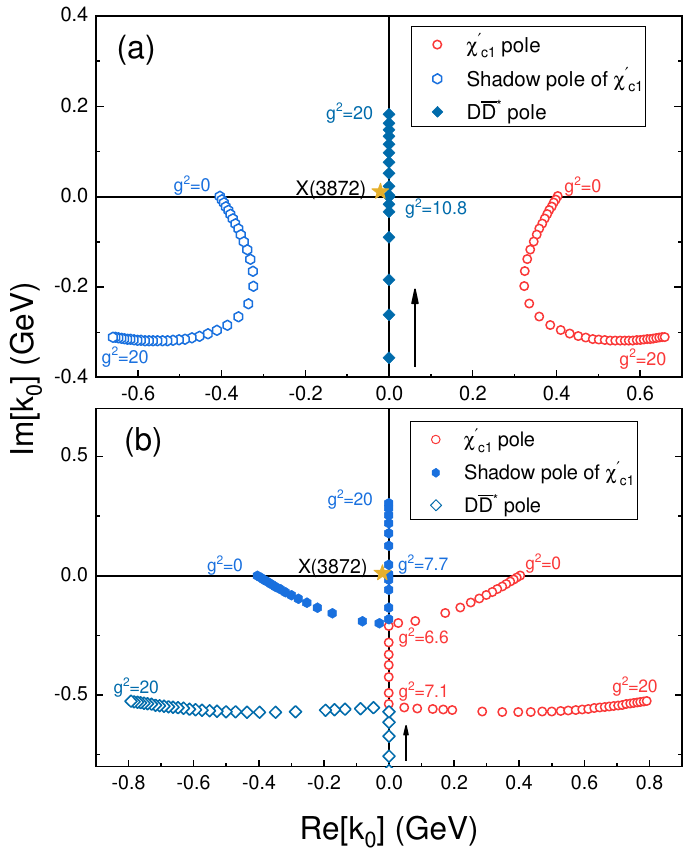} 
    \caption{\label{fig1} The pole evolutions of the single-channel $D\bar{D}^{*} \to D\bar{D}^{*}$ scattering in the coupled-channel dynamics. Here, the complex plane refers to the momentum $k_0$.}
\end{figure}

We first focus on the situation with the strict isospin symmetry, in which the two-channel calculation in Eq.~(\ref{eq11}) is reduced to a single-channel one. The pole trajectories of the $D\bar{D}^{*}$  scattering in the momentum $k_0$-plane when increasing the coupling strength $g$ are presented in Fig.~\ref{fig1}.  Here, $k_0$ is defined by $k_0=\sqrt{2\mu(E-m_D-m_{D^*})}$. $k_0$ with positive and negative imaginary parts correspond to the first and second Riemann sheets of the $T$ matrix, respectively. The  bare mass $M_0=3.96$ GeV is adopted according to the quark model estimations \cite{Godfrey:1985xj,Barnes:2005pb}. Fig.~\ref{fig1}($a$) and ($b$) with no contact interaction ($C_t=0$ $\text{GeV}^{-2}$)   corresponds to the so-called weak coupling mode with a typical value $\alpha=1.0$ GeV and in the strong coupling mode with $\alpha=1.4$ GeV, respectively, which indicates a smaller and larger overlap range between the wave functions of  the $c\bar{c}$ core and $D\bar{D}^{*}$ channel.

In the weak coupling mode, there appear three poles of the $T$ matrix in the complex plane, although  only the $s$-channel interaction $\mathcal{V}$ is  considered.  It is easy to identify the nature of these three poles by setting $g^2 \to 0$. The red, blue and green poles correspond to the bare $\chi_{c1}^{\prime}$ resonance, the shadow pole associated with $\chi_{c1}^{\prime}$ resonance  and a virtual state of $D\bar{D}^*$ at the infinity, respectively. With gradually increasing $g^2$,  the pole of $\chi_{c1}^{\prime}$ resonance moves first towards and then away from the threshold of $D\bar{D}^*$, while the $D\bar{D}^*$ virtual state continuously moves towards the threshold and finally crosses the unitary branch cut of $D\bar{D}^*$ and becomes a bound state, which corresponds to the observed $X(3872)$ structure. For the strong coupling mode shown in Fig.~\ref{fig1}($b$), the situation is completely different. When $g^2$ increases to a critical value, the $\chi_{c1}^{\prime}$ pole and its shadow pole will meet at the same point as virtual states and then continue to move along the negative imaginary axis of the $k_0$-plane in opposite directions. Subsequently, the $\chi_{c1}^{\prime}$ pole will meet the $D\bar{D}^*$ virtual state pole arising from the infinity and then they evolve into a pair of resonance and anti-resonance.  In contrast, the shadow pole associated with $\chi_{c1}^{\prime}$ enters the first Riemann sheet and is related to a $X(3872)$ state. 

{It is important to highlight that this pole separation behavior associated with the $\chi_{c1}^{\prime}$ state is directly attributed to the tiny imaginary part $i\epsilon$ of the effective potential 
\begin{align}
 \mathcal{V} \propto (m_{th}+k^2_0/(2\mu)-M_0+i\epsilon)^{-1}.    
\end{align}
Thus, such a phenomenon is still universal when including the intrinsic  $D\bar{D}^*$ dynamics, as shown in Supplemental Material (SM) \cite{Suppl}. }  It can be concluded that an intrinsic attractive force between $D$ and $\bar{D}^*$ prompts the pole evolution behaviors  to favor the weak coupling mode and a repulsive force causes poles to evolve in favor of the strong coupling mode.
To sum up, whatever the dynamics origin of $X(3872)$ is,  
\emph{$X(3872)$ does not originate from the mass shift of the bare $\chi_{c1}^{\prime}$ resonance pole in the coupled-channel dynamics.}

\begin{figure}
    \centering
    \includegraphics[width=0.49\textwidth]{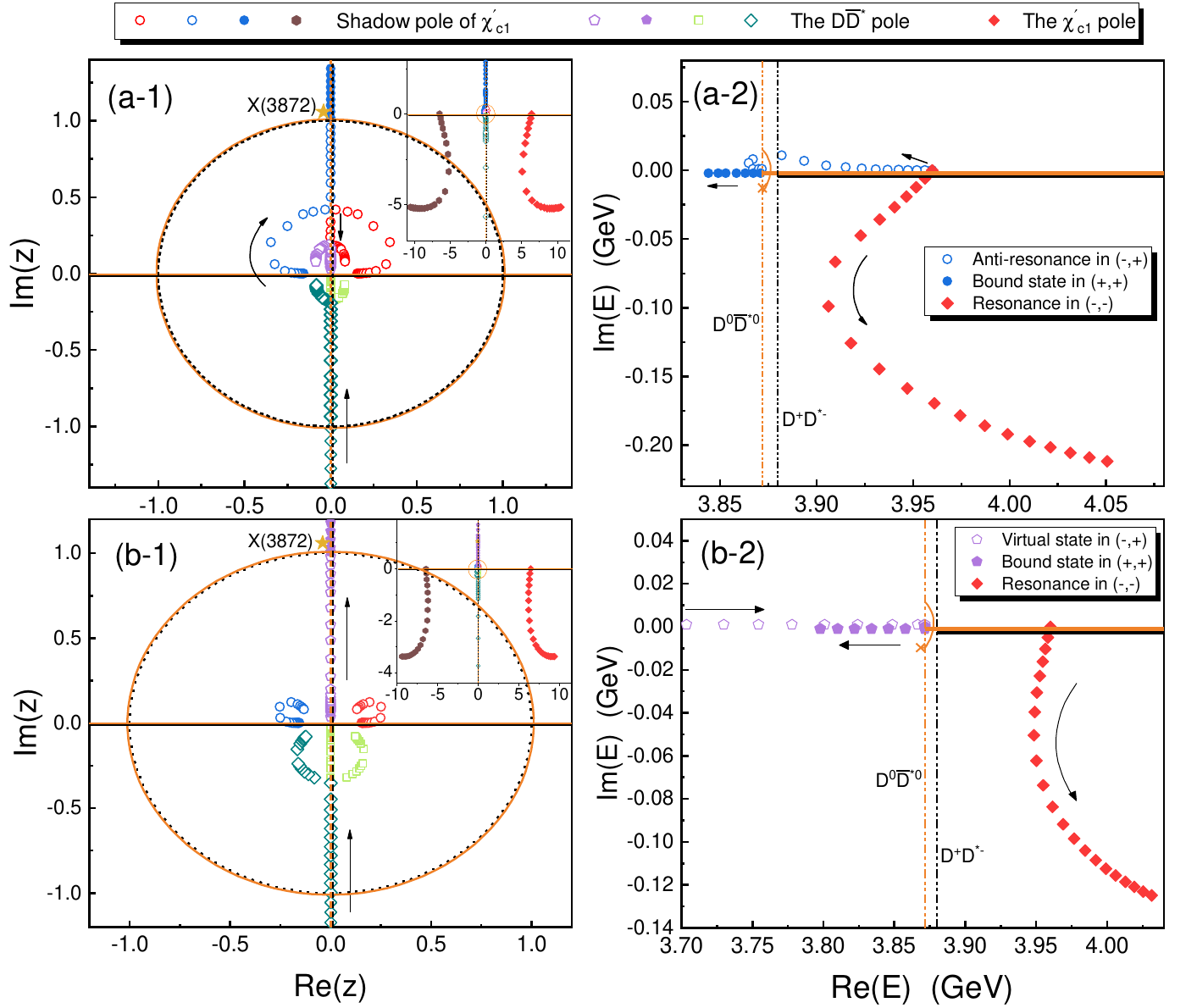} 
    \caption{\label{fig2}  The pole evolutions of the two-channel $D\bar{D}^{*} \to D\bar{D}^{*}$ scattering (isospin violation) in the coupled-channel dynamics. Left and right figures are related to the uniformization $z$-plane and the corresponding energy $E$-plane. It is worth emphasizing that the solid and hollow blue (purple) points refer to the same pole.}
\end{figure}

For $X(3872)$, the isospin breaking effect is significant since the mass difference between the neutral and charged channels is up to  8 MeV \cite{ParticleDataGroup:2022pth}. Thus, we further study the pole evolution of $D\bar{D}^* \to D\bar{D}^*$ in the two-channel scattering, which is presented in Fig.~\ref{fig2}. Here, in order to make the $T$-matrix a single-valued function in the locally flat surface, a uniformization variable $z$ is introduced to achieve a mapping from energy $E$ to $z$ \cite{Yamada:2020rpd,Yamada:2021azg,Meng:2022wgl}, whose definition is summarized in the SM \cite{Suppl}.
In the $z/E$-plane of Fig.~\ref{fig2}, the orange and black solid lines stand for the unitary branch cuts related to the $[D^0\bar{D}^{*0}]$ and $[D^+\bar{D}^{*-}]$, respectively. The dashed lines represent the real axis of the $E$-plane unoccupied by the branch cut. The four Riemann sheets $(+,+)$, $(-,+)$, $(+,-)$ and $(-,-)$ for the two-channel system correspond to the upper half $z$-plane outside and inside the circle, and the lower half $z$-plane inside and  outside the circle, respectively, where $+$ and $-$ represent the first and the second Riemann sheets, respectively.

Since the impact of the intrinsic $D\bar{D}^*$ dynamics on the pole evolution in the two-channel case is very similar to the lesson learned from the single-channel analysis, here we only show the two-channel calculations  with $C_t=C_t^{\prime}=0$ $\text{GeV}^{-2}$. In Fig.~\ref{fig2}($a$-1) with a typical value $\alpha=1.0$ GeV, a large number of new poles appear due to the isospin symmetry violation, and {it can be found that the evolution behaviors of the dressed $\chi_{c1}^{\prime}$ resonance and its shadow pole in the $(-,-)$ sheet are very similar to those of the weak coupling mode in the single-channel case.} However, the $X(3872)$ pole no longer arises from the virtual $D\bar{D}^*$ state in the $(-,-)$ sheet. Instead, it arises from the shadow resonance of $\chi_{c1}^{\prime}$ in the $(-,+)$ sheet. In Fig.~\ref{fig2}($a$-2), we show the corresponding pole trajectories  in the complex energy plane. It can be seen that the dressed $\chi_{c1}^{\prime}$ resonance is always higher than the $[D^+\bar{D}^{*-}]$ threshold. Its shadow pole in the $(-,+)$ sheet first moves into the energy region below the $[D^0\bar{D}^{*0}]$ threshold and then returns and crosses the unitary cut of $[D^0\bar{D}^{*0}]$ into the $(+,+)$ sheet and becomes a bound state $X(3872)$. This trajectory is still maintained when employing a stronger coupling mode such as $\alpha=$ 1.4 GeV. However, when adopting a weaker coupling mode ($\alpha=$ 0.8 GeV), as presented in  Fig.~\ref{fig2}($b$-1) and ($b$-2), the shadow pole of $\chi_{c1}^{\prime}$ in the $(-,+)$ sheet will not be pulled to the imaginary axis of the $z$-plane and the $X(3872)$ originates from a $D\bar{D}^*$ virtual state pole in the $(-,+)$ sheet.

 { \it A novel approach to reveal the nature of $X(3872)$ }---Recently, the LHCb and BESIII Collaborations extracted the pole position of $X(3872)$ by analyzing the line shape, whose imaginary parts are $-(130^{+320}_{-180})i$ keV \cite{LHCb:2020xds}  and $-(190^{+206}_{-161})i$ keV \cite{BESIII:2023hml}, respectively. Both of the center absolute values are obviously larger than $-(24\pm1)i$ keV  \cite{LHCb:2021vvq,LHCb:2021auc} of $T_{cc}^+$. The pole width of $T_{cc}^+$ has been reproduced well in theoretical calculations, attributed mainly to its three-body $DD\pi$ decay \cite{Meng:2021jnw,Du:2021zzh,Lin:2022wmj,Cheng:2022qcm,Ling:2021bir}.  However, the $D\bar{D}\pi$ half decay width of $X(3872)$ from the three-body cut of the OPE potential is estimated to be only $15\sim34$ keV in Refs. \cite{Lin:2022wmj,Cheng:2022qcm}, and the reason for the large pole width of $X(3872)$ remains unclear in the current theoretical studies.

\begin{figure}
    \centering
    \includegraphics[width=0.48\textwidth]{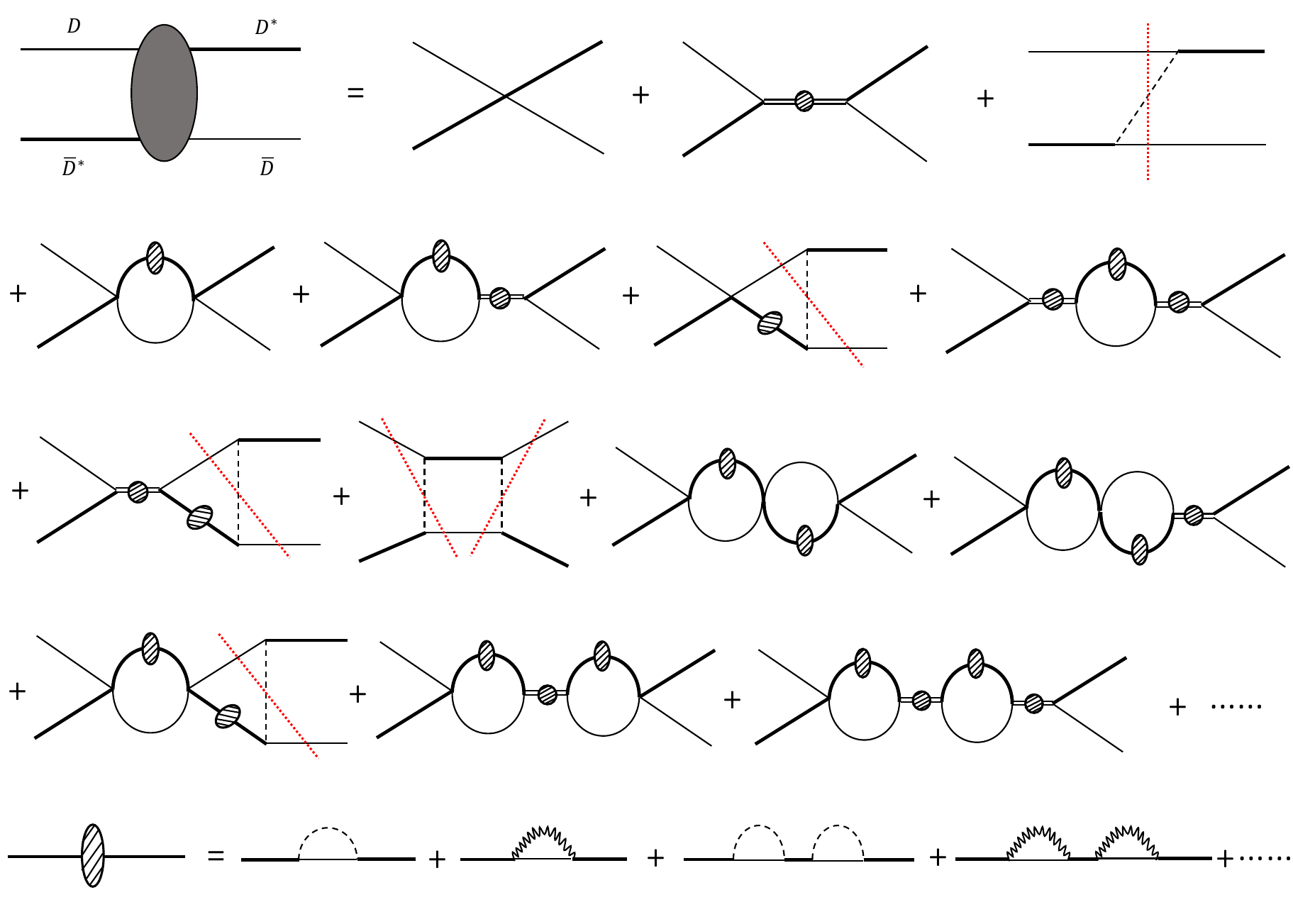} 
    \caption{\label{fig3}  { The full Feynman diagram of the $D\bar{D}^* \to D\bar{D}^*$ scattering } in the coupled-channel dynamics. The self-energy diagram from the non-open-charm decays for the bare $\chi_{c1}^{\prime}$ is marked by shadow circle.}
\end{figure}

{ Here, we perceive that the coupled-channel dynamics plays a key role in the large pole width of $X(3872)$. This idea is supported by the fact that the experimental total width of the ground state charmonium $\chi_{c1}(1P)$ without open-charm decay channels, can reach $0.88\pm0.05$ MeV \cite{ParticleDataGroup:2022pth}.  Hence, it is reasonable to expect that the non-open-charm decay widths of the bare $\chi_{c1}^{\prime}$ state are of a similar order of magnitude.
We employ a quark-model-independent scheme to estimate the decay behaviors of the bare $\chi_{c1}^{\prime}$ state. The relevant details can be found in the SM \cite{Suppl}. The coupling amplitude $\mathcal{V}_{0j}(\textbf{q})$ and $\mathcal{V}_{i0}(\textbf{q}^{\prime})$ in Eq.~(\ref{eq8}) can also be obtained in the quark pair creation (QPC) model without extra parameter dependence. Now, the $s$-channel effective potential should be modified as
\begin{align}
    \mathcal{V}_{ij}(\textbf{q},\textbf{q}^{\prime})=\frac{\mathcal{V}_{i0}(\textbf{q}^{\prime})\mathcal{V}_{0j}(\textbf{q})}{E-M_0+i\epsilon} \to \frac{\mathcal{V}_{i0}(\textbf{q}^{\prime})\mathcal{V}_{0j}(\textbf{q})e^{-\lambda^2(\textbf{q}^2+\textbf{q}^{\prime 2})}}{E-M_0+i\frac{1}{2}(\Gamma_a+\Gamma_b)}, 
\end{align}
where $\Gamma_a=3488$ keV and $\Gamma_b=220$ keV  correspond to the light hadron decays and radiative decays of the bare $\chi_{c1}^{\prime}$, respectively. 
According to suggestions from Refs. \cite{Morel:2002vk,Ortega:2016mms,Yang:2021tvc}, an extra cutoff parameter $\lambda$ is required to adjust the large momentum suppression in $\mathcal{V}_{i0}$ from the input of the QPC amplitude. Additionally, we also simultaneously take into account the  three-body $D\bar{D}\pi$ threshold effect from the OPE interaction \cite{Lin:2022wmj} and the dynamical width of $\bar{D}^*$ from its strong decay $\bar{D}\pi$ and electromagnetic decay $\bar{D}\gamma$. The complete scattering amplitudes of $D\bar{D}^* \to D\bar{D}^*$ involving these three potential mechanisms contributing to the pole width are presented in Fig.~\ref{fig3}.  }

{ The inclusion of these decay dynamics makes the system Hamiltonian no longer Hermitian. In this situation, a more convenient approach to extract the pole positions with the full scattering amplitudes in Fig.~\ref{fig3} is the complex scaling method (CSM) \cite{Aguilar:1971ve,Balslev:1971vb,Myo:2014ypa}. 
In our previous studies on $T_{cc}^+$ with ChEFT \cite{Lin:2022wmj}, the pole width is not sensitive to the cutoff $\Lambda$ within a large range $\Lambda=0.4-1.0$  GeV, and then the $\Lambda$ cutoff dependence can be safely cancelled by renormalizing the contact interaction. Thus, the $\Lambda$ is fixed to $0.5$ GeV and the undetermined parameters are $\lambda$, $C_t$ and $C_t^{\prime}$. The last two can be related to two new contact constants $V_{ct}^{I=0}$ and $V_{ct}^{I=1}$ (see SM for their definitions \cite{Suppl}). 
First, we ignore the isospin violation of the contact interaction, i.e., $V_{ct}^{I=1}=0$. The solved observable poles and their corresponding component possibilities are listed in Table \ref{tab:1}.  For different cutoff values $\lambda$, we adjust $V_{ct}^{I=0}$ in a one-to-one manner to match the small binding energy of $X(3872)$. When only the width $\Gamma_a+\Gamma_b$ from the bare state is involved, whose contribution can be transferred into other poles via the coupled-channel mechanism, the pole width of $X(3872)$ can be significantly enlarged to the order of $\mathcal{O}(10^{-1})$ MeV, which highlights the crucial role of coupled-channel dynamics.  }

{ As a notable prediction, a higher resonance pole is found around 4 GeV as shown in Table \ref{tab:1}, which corresponds either to the genuine dressed charmonium $\chi_{c1}^{\prime}$ resonance as argued in Refs. \cite{Kalashnikova:2005ui,Coito:2012vf,Cincioglu:2016fkm,Zhou:2017dwj,Giacosa:2019zxw,Wang:2023ovj,Li:2024pfg} or to a distorted $D\bar{D}^*$ resonance. In the latter case, the $\chi_{c1}^{\prime}$ resonance evolves into a virtual pole far from the physical area in the $(-,-)$ sheet.  Interestingly, we find that the pole width of $X(3872)$ and the properties of the new resonance are highly sensitive to the cutoff $\lambda$, with its variation from small to large values directly reflecting the transition from strong to weak coupling modes, as introduced in the previous section. For instance, for $\lambda=0.5~\text{GeV}^{-1}$, the $X(3872)$ pole originates from the shadow resonance of $\chi_{c1}^{\prime}$ in the $(-,+)$ sheet, whereas for $\lambda=2.5~\text{GeV}^{-1}$, it corresponds to the $D\bar{D}^*$ pole in the same sheet.  { Additionally, as shown in Table~\ref{tab:1}, poles with a larger $c\bar{c}$ component typically correspond to strong coupling mode. Conversely, poles with a dominant $D\bar{D}^*$ component tend to correspond to a weak coupling mode.} The results assuming an evident isospin violation of the contact interaction are summarized in the SM \cite{Suppl}, in which the similar conclusions are obtained. This remarkable feature inspires a new idea: the characteristic physical quantity of $X(3872)$---pole width or the predicted new resonance can be treated as a novel indicator to convincingly reveal the nature of $X(3872)$. { In particular, a very narrow pole width of $X(3872)$ on the order of tens of keV, similar to that of $T_{cc}(3875)$, strongly implies a dominant molecular composition of $X(3872)$ and the origin of the $D\bar{D}^*$ pole. On the other hand, an imaginary part of the $X(3872)$ pole in the hundreds of keV to MeV would suggest a significant compact $c\bar{c}$ admixture and favor the origin of the shadow pole of $\chi_{c1}^{\prime}$. Unfortunately, due to current experimental precision limitations, the imaginary part of the $X(3872)$ pole is still not well determined. Future high-precision measurements could serve as a powerful discriminator for the internal structure of $X(3872)$. }

Thanks to the recent LHCb observation of $\chi_{c1}(4010)$, determined with  an exceptionally high statistical significance ($16\sigma$) through a new amplitude analysis {  involving simultaneous fits to two $C$-parity-related decays $B^+ \to D^+D^{*-}K^+$ and $B^+ \to D^-D^{*+}K^+$} \cite{LHCb:2024vfz}, our prediction is strongly validated. { Here, it is worth emphasizing that the $B \to D\bar{D}^*K$ is a complicated decay process and the reported $\chi_{c1}(4010)$ state is still single observation, which requires further confirmation by other experiments. If we assume that the $\chi_{c1}(4010)$ state exists, } for the following three reasons, the identification of $\chi_{c1}(4010)$ as our predicted resonance is highly reasonable. First, the $\chi_{c1}(4010)$ was observed in the invariant mass spectrum of $D^{*\pm}D^{\mp}$, the dominant decay mode of our predicted resonance. Second, although its mass lies close to the threshold of $D^*\bar{D}^*$, the existence of an $S$-wave $D^*\bar{D}^*$ structure with $J^{PC}=1^{++}$ is forbidden due to the odd $C$-parity constraint for particle-antiparticle systems. Third, higher $P$-wave charmonium excitations are significantly beyond the mass range of $\chi_{c1}(4010)$, for example the $\chi_{c1}(3P)$, which was widely predicted to reside around 4.3 GeV \cite{Godfrey:1985xj,Barnes:2005pb}. These factors make alternative explanations for $\chi_{c1}(4010)$ highly unlikely. We match our predicted resonance to the newly observed $\chi_{c1}(4010)$ and present the experimental constraints incorporated in two-dimension Gaussian distribution function on parameter $\lambda$, with detailed results shown in the SM \cite{Suppl}.  Although current experimental information on the pole width of $X(3872)$ does not distinguish its two pole origins, the $\chi_{c1}(4010)$ imposes strong constraints on the coupling modes of the coupled-channel dynamics. Ultimately, we conclude with over $99.7\%$ confidence level that $X(3872)$ originates from the $D\bar{D}^*$ pole in the $(-,+)$ Riemann sheet.   }

\begin{table*}[t]
\renewcommand{\arraystretch}{2}
    \centering
\caption{ The poles and component possibilities\footnote{ Here, the component possibility $\mathcal{P}$ is defined as $\mathcal{P}_i=( \phi_i |\phi_i )$ with the wave function of $i$-th channel $\phi_i$, which roughly reflects the ratio of the $i$-th channel.  For the non-Hermitian system, the wave function normalization is defined by c-product \cite{Romo:1968tcz,Lin:2023ihj}, which means 
  $ ( \phi| \phi )=\sum_i\int\frac{d\bm{p}^3}{(2\pi)^3}e^{-3i\theta}\phi_i(\tilde{\bm p})^2=1,$ 
where $\theta$ is the rotation angle in complex scaling method, and subscript $i$ represents the involved channels. } of $X(3872)$ and a new resonance in the complete coupled-channel dynamics. The pole of $X(3872)$ is given by the binding energy relative to the $D^0\bar{D}^{*0}$ threshold. The $\chi$ and $M$ superscripts denote the dressed $\chi_{c1}^{\prime}$ resonance and the distorted  $D\bar{D}^*$ resonance in the $(-,-)$ sheet, respectively. Here, $V_{ct}^{I=1}=0$ $\text{GeV}^{-2}$ and all pole positions are in units of MeV.  
{It is important to emphasize that, while the real part of the experimentally extracted pole for $X(3872)$ is slightly positive, it remains on the first Riemann sheet. In this analysis, we compare the experimental results with a theoretical bound-state pole whose real part is slightly negative. This difference does not compromise the validity of the comparison, as such a pole can be continuously shifted above the threshold, staying on the same Riemann sheet, through a slight reduction in the interaction strength.} }
\setlength{\tabcolsep}{2.0mm}{
    \begin{tabular}{c|c|c|cccccccc}
    \hline
     \hline
     \multicolumn{3}{c|}{$\lambda$ ($V_{ct}^{I=0}$ ) ~~$\text{GeV}^{-1}$ $(\text{GeV}^{-2})$ }  & 0.5 (96.6)  & 1.0 (22.5) & 1.25 (11.2)  &1.5 (3.7) & 2.5 (-13.1)   \\
    \hline
    Without $\Gamma_a+\Gamma_b$ & $X(3872)$  & Pole &   -0.086-0.003i &   -0.140-0.024i &   -0.097-0.027i  &  -0.089-0.029i & -0.075-0.030i \\
    \hline
     \multirow{8}{*}{With $\Gamma_a+\Gamma_b$}     & \multirow{4}{*}{$X(3872)$} & Pole & -0.059-1.36i  &  -0.060-0.293i &  -0.060-0.164i &  -0.070-0.119i  &  -0.071-0.065i    \\
    \cline{3-8}
    & & $\mathcal{P}_{D^0\bar{D}^{*0}}$ & 0.145-0.015i &  0.748-0.136i &   0.858-0.078i &  0.895-0.047i  &  0.939-0.013i    \\
    \cline{3-8}
   & & $\mathcal{P}_{D^+\bar{D}^{*-}}$ & 0.110+0.002i  &  0.092+0.049i &  0.065+0.035i &  0.056+0.024i   &  0.043+0.009i    \\
    \cline{3-8}
    & & $\mathcal{P}_{\chi_{c1}^{\prime}}$ & 0.745+0.013i  &  0.160+0.087i &  0.077+0.043i &  0.049+0.023i  &  0.018+0.004i    \\
    \cline{2-8}
      & \multirow{4}{*}{New resonance} & Pole   & 4150-141i$^{M}$  &  4063-129i$^{M}$  &  4025-92i$^{\chi}$ &  4004-57i$^{\chi}$ &  3977-8i$^{\chi}$    \\
       \cline{3-8}
       & & $\mathcal{P}_{D^0\bar{D}^{*0}}$ & 0.328+0.048i  &  0.278-0.191i &  0.203-0.255i &  0.104-0.209i  &  0.088-0.040i   \\
    \cline{3-8}
    & & $\mathcal{P}_{D^+\bar{D}^{*-}}$ & 0.324+0.007i  &  0.265-0.219i &  0.187-0.286i &  0.086-0.234i  &  0.092-0.062i    \\
    \cline{3-8}
   & & $\mathcal{P}_{\chi_{c1}^{\prime}}$ & 0.348-0.055i  &  0.457+0.410i &  0.610+0.541i &  0.810+0.443i  &  0.820+0.102i    \\
       \hline
     \multicolumn{2}{c|}{BESIII \cite{BESIII:2023hml} } & Pole & \multicolumn{3}{c}{$(0.0068^{+0.1655}_{-0.1700})-(0.190^{+0.206}_{-0.161})i$} \\
       \hline
       \multicolumn{2}{c|}{LHCb \cite{LHCb:2020xds} }  & Pole &  \multicolumn{3}{c}{$(0.06^{+0.16}_{-0.16})-(0.13^{+0.32}_{-0.18})i $}\\
    \hline
     \hline
    \end{tabular}
    }
    \label{tab:1}
\end{table*}

{\it Summary.}---We have studied the pole origin of $X(3872)$ with the coupled-channel dynamics and definitely demonstrated that the $X(3872)$ does not originate from the mass shift of the bare $\chi_{c1}^{\prime}$ resonance pole. Accordingly, 
it stems from either the $D\bar{D}^*$ virtual state pole in the weak-coupling mode or the shadow pole associated with $\chi_{c1}^{\prime}$ in the strong-coupling mode. Furthermore, we focused on the pole width of $X(3872)$ by considering the complete scattering dynamics, incorporating three key dynamical mechanisms. Our findings indicate that  the coupled-channel dynamics will result in an apparently larger pole width of $X(3872)$. 
{ More intriguingly, we have established a quantitative connection among the dynamical origin of $X(3872)$, its pole width and the properties of the predicted new resonance. This underscores the importance of precisely measuring the pole width of $X(3872)$ and the characteristics of the higher resonance in fully resolving the puzzle of $X(3872)$. By matching the predicted resonance to the newly observed $\chi_{c1}(4010)$, we conclusively identify the $D\bar{D}^*$ pole as the origin of $X(3872)$ with overwhelming confidence. }

\begin{acknowledgments}
This work is supported by the National Science Foundation of China under Grants No. 11975033, No. 12070131001 and No. 12147168. This project was also funded by the Deutsche Forschungsgemeinschaft (DFG,
German Research Foundation, Project ID 196253076-TRR 110). J.Z.W. is also supported by the National Postdoctoral Program for Innovative Talent.

{\it Note added.}---About one month after this work was posted on arXiv, we notice that the LHCb Collaboration released the amplitude analysis results of $B^+ \to D^{*\pm}D^{\mp}K^+$ and observed a new charmoniumlike state $\chi_{c1}(4010)$ with quantum number $J^{PC}=1^{++}$ decaying into $D^{*\pm}D^{\mp}$, whose mass and width are fitted to be $4012.5^{+3.6+4.1}_{-2.9-3.7}$ and $62.7^{+7.0+6.4}_{-6.4-6.6}$ MeV \cite{LHCb:2024vfz}, respectively. The newly observed $\chi_{c1}(4010)$ surprisingly coincides with our predicted higher resonance around 4.0 GeV, which strongly supports the key role of the coupled channel dynamics in the formation of $X(3872)$. 
\end{acknowledgments}

\bibliography{ref}

\begin{onecolumngrid}
\appendix
\newpage

\section{The pole trajectories of the $D\bar{D}^{*}$  scattering when including the intrinsic  $D\bar{D}^*$ dynamics}

In Fig.~\ref{fig4}, we show the pole trajectories of the single-channel $D\bar{D}^{*}$  scattering with increasing the coupling strength $g$  when including the intrinsic  $D\bar{D}^*$ dynamics. The subfigures $a$-1, $a$-2, $b$-1, and $b$-2 represent the cases with an attractive contact interaction ($C_t=-10$) $\text{GeV}^{-2}$ and a typical coupling $\alpha=1.0$ GeV, a repulsive contact interaction ($C_t=10$ $\text{GeV}^{-2}$) and  a typical coupling $\alpha=1.0$ GeV, an attractive contact interaction ($C_t=-10$) $\text{GeV}^{-2}$ and a typical coupling $\alpha=1.4$ GeV, and a repulsive contact interaction ($C_t=10$ $\text{GeV}^{-2}$) and a typical coupling $\alpha=1.4$ GeV, respectively.  

It can be seen that an intrinsic attractive force between $D$ and $\bar{D}^*$ prompts the pole evolution behaviors  to favor the weak coupling mode and a repulsive force causes poles to evolve in favor of the strong coupling mode.
Interestingly, due to the influence of the repulsive interaction, the $\chi_{c1}^{\prime}$ pole might continuously emerge as a virtual state by a strong enough threshold coupling.

\begin{figure*}[h]
\centerline{\includegraphics[width=16.0cm]{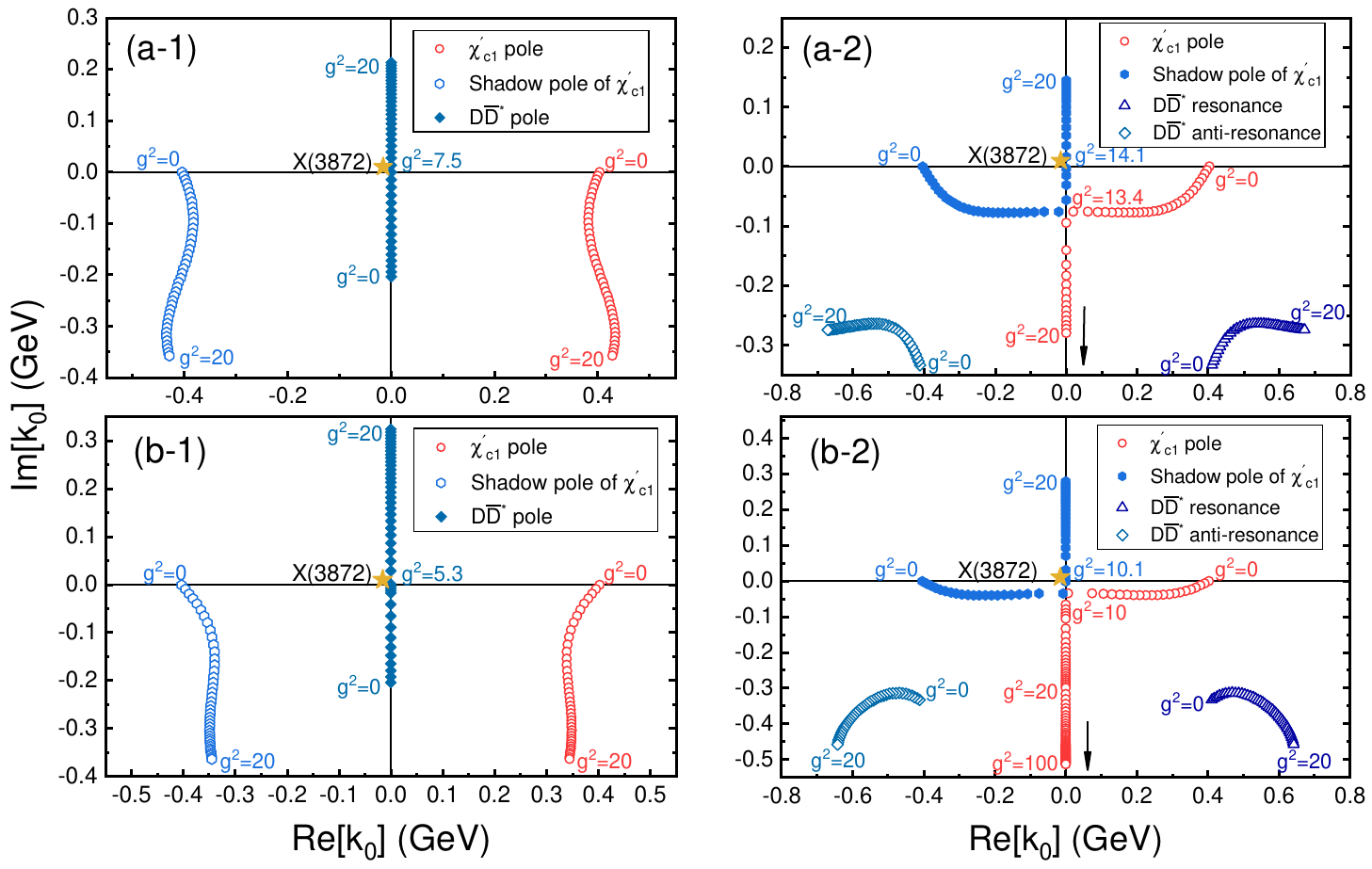}}
	\caption{ The pole evolutions of the single-channel $D\bar{D}^{*} \to D\bar{D}^{*}$ scattering (no isospin violation) in the coupled-channel dynamics. Here, the complex plane refers to the momentum $k_0$. The subfigure   (a-1), (a-2), (b-1) and (b-2) correspond to the $(C_t~(\text{GeV}^{-2}), \alpha~(\text{GeV}))$=(-10, 1.0), (+10, 1.0), (-10, 1.4) and (+10, 1.4), respectively. 
 \label{fig4} }
\end{figure*}

\section{  The experimental constraints on the pole positions of $X(3872)$ and the predicted new resonance  } 

{We solve the poles of $X(3872)$ and the accompanying new resonance for different values of $\lambda=2.50, 1.70, 1.60, 1.50, 1.40, 1.25, 1.00, 0.90$ GeV$^{-1}$. Additionally, we investigate the impact of the bare mass $m_0$ of $\chi_{c1}^{\prime}$ on the two pole positions. To present the experimental constraints on the parameter space of $\lambda$, we use a two-dimensional Gaussian function to reflect the experimental pole distribution possibilities of $X(3872)$ and the predicted new resonance in the complex energy plane. The variances $\sigma_i$ represent the experimental errors associated with the extracted pole positions. In Fig. \ref{Appen:X}, we show the experimental constraints on the pole position of $X(3872)$ based on data from BESIII and LHCb \cite{BESIII:2023hml,LHCb:2020xds}, where $\delta E$ is the binding energy. The red and blue points represent the theoretical solutions of $X(3872)$ in the strong and weak coupling modes, respectively. Here, the dots with the same pattern refer to theoretical results with the same $\lambda$ but different coupling constant $V_{ct}^{I=0}$, whose impact is almost reflected in the binding energy of $X(3872)$ and is therefore only  shown in the case with $m_0=3.96$ GeV. It can be seen that the entire parameter space of $\lambda$, from small to large, is allowed by the experimental information of $X(3872)$, meaning that current measurement cannot distinguish between the two possible pole origins of $X(3872)$. }

{By matching newly observed $\chi_{c1}(4010)$ \cite{LHCb:2024vfz} to the predicted new resonance, the experimental constraints on its pole position are shown in Fig. \ref{Appen:4010}. For the case with $m_0=3.96$ GeV, it can be seen that the solved poles with the same $\lambda$ but different $V_{ct}^{I=0}$ are completely overlapping, so the new resonance poles have very little dependence on the $V_{ct}^{I=0}$.  Remarkably, the LHCb measurement of $\chi_{c1}(4010)$ actually provides strong constraints on the parameter $\lambda$. The solved poles of the new resonance near $\lambda=1.60$ GeV$^{-1}$ consistently fall within the $99.7\%$ confidence level for different bare masses $m_0=3.95, 3.96, 3.97$ GeV, which correspond to the weak-coupling mode. Consequently, the red point solutions, which correspond to the pole origin of $X(3872)$ as the shadow pole of $\chi_{c1}^{\prime}$, can essentially be  excluded.  }

\begin{figure*}[t]
    \centering
    \begin{tabular}{ccc} 
        \includegraphics[width=0.33\textwidth]{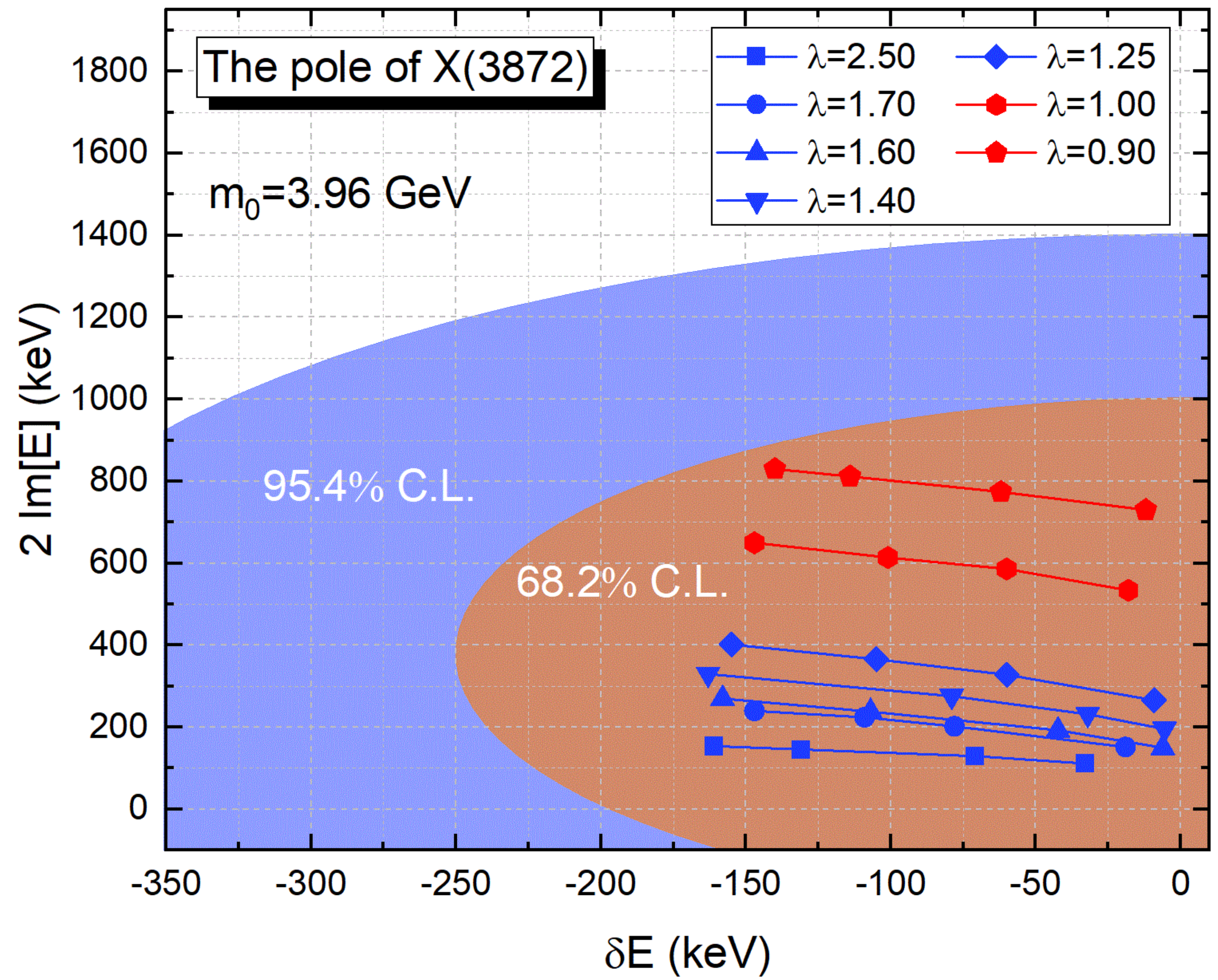} & 
        \includegraphics[width=0.31\textwidth]{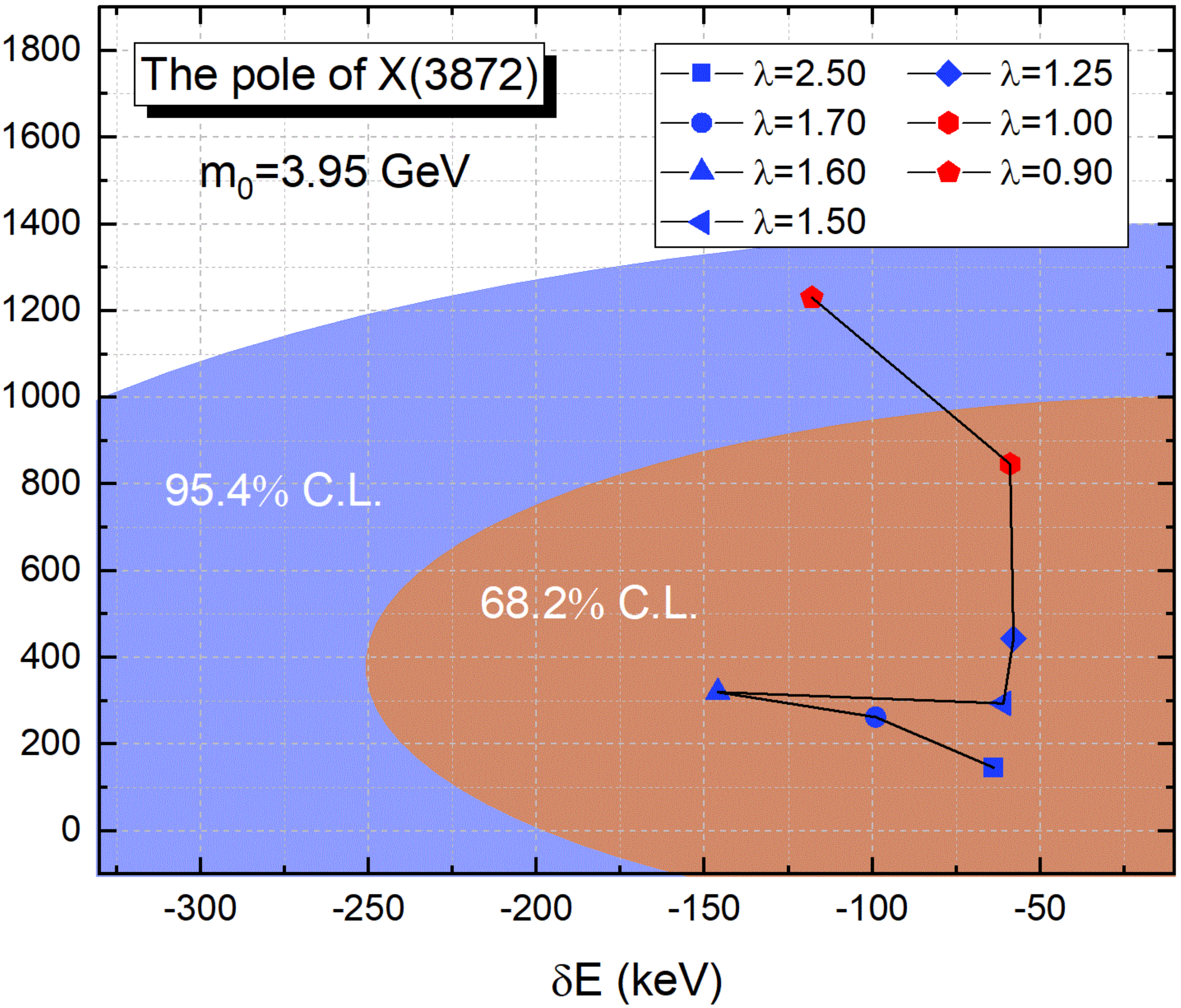} & 
        \includegraphics[width=0.31\textwidth]{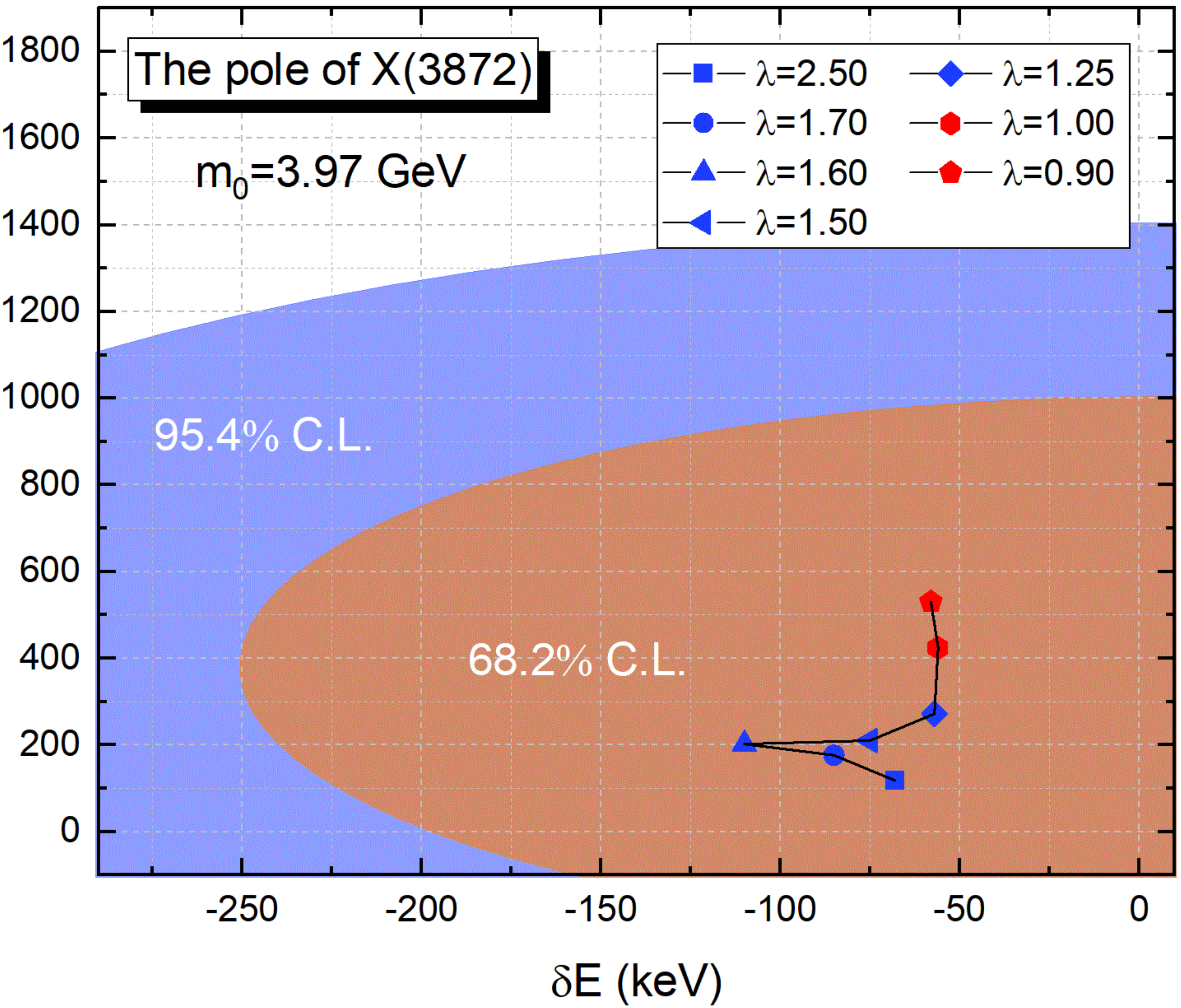} 
    \end{tabular}
    \caption{  The experimental constraints on the pole position of $X(3872)$. The shaded region represents the confidence level for the pole distribution possibility of $X(3872)$ based on the experimental data. The solid dots represent the theoretical pole positions of $X(3872)$, with red and blue representing the solutions in the strong and weak coupling modes, respectively. The subfigures, from left to right, correspond to the bare mass values $m_0=3.96, 3.95, 3.97$ GeV for the $\chi_{c1}^{\prime}$. Here, the dots with the same pattern refer to theoretical results with the same $\lambda$ but different coupling constant $V_{ct}^{I=0}$, whose impact is almost reflected in the binding energy of $X(3872)$ and is therefore only  shown in the subfigure on the left. \label{Appen:X}}
\end{figure*}

\begin{figure*}[h]
\centerline{\includegraphics[width=10.0cm]{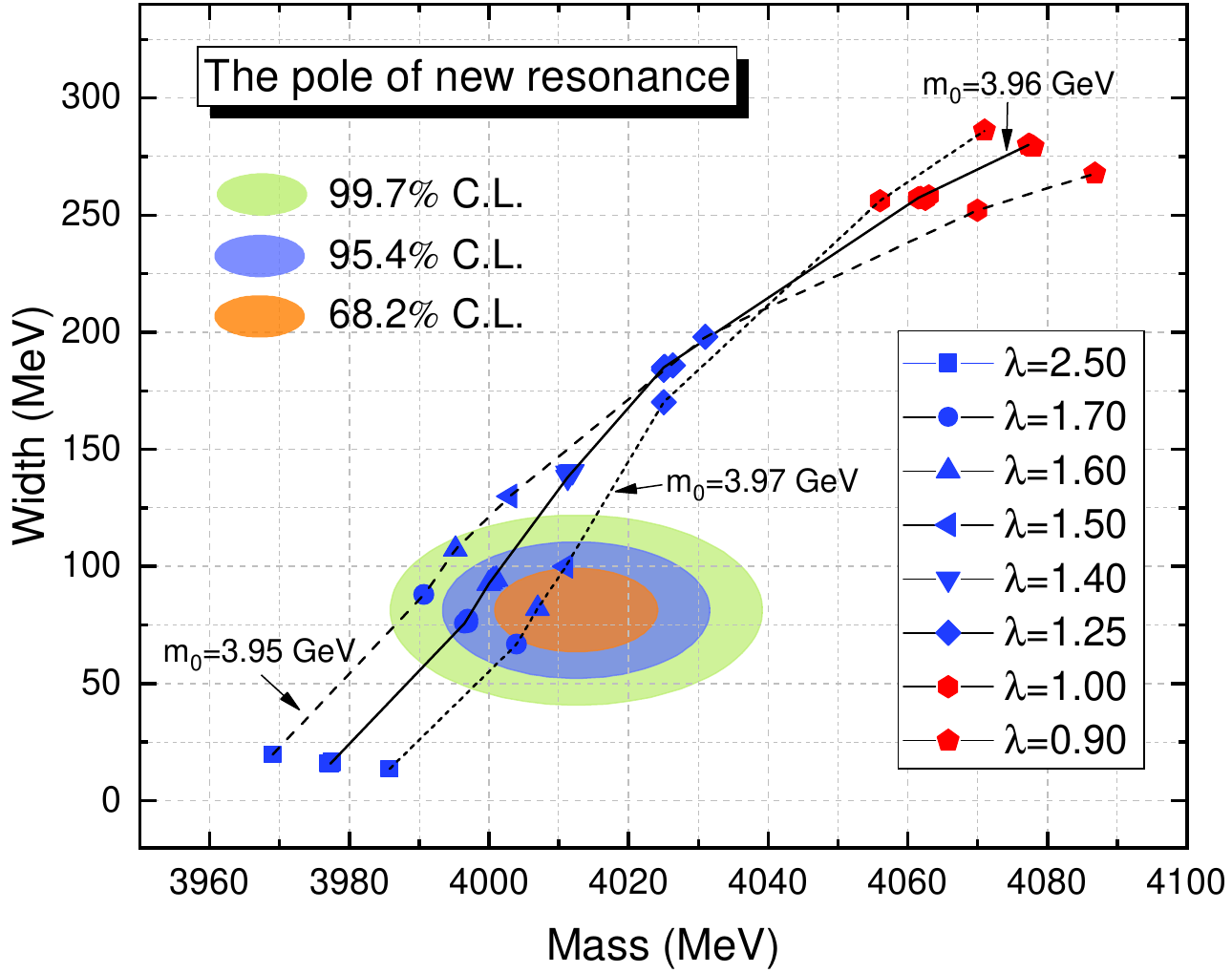}}
	\caption{  The experimental constraints on the pole position of the predicted new resonance. The shaded region represents the confidence level for the pole distribution possibility of $\chi_{c1}(4010)$ based on the LHCb data. The solid dots represent the theoretical pole positions of the new resonance, with red and blue representing the solutions in the strong and weak coupling modes, respectively. \label{Appen:4010} }
\end{figure*}

{Meanwhile, we clearly identify the $\chi_{c1}(4010)$ as the genuine dressed $P$-wave charmonium excitation, which is very important for understanding the messy situation associated with the $2P$-wave charmonium spectroscopy. Very intriguingly, in our explanation, the mass of the $J^{PC}=1^{++}$ charmonium state ($\chi_{c1}(4010)$) is obviously larger than that of the $J^{PC}=2^{++}$ charmonium state ($\chi_{c2}(3930)$). 
In fact, this abnormal phenomenon can be naturally understood in the significant coupled-channel dynamics between the $\chi_{c1}(2P)$ and the $D\bar{D}^*$ continuum, causing the mass of the $\chi_{c1}(2P)$ resonance pole to increase rather than decrease as typically expected,  as presented in Fig. 2 of the main text.  This implies that, although the $X(3872)$ corresponds to a $D\bar{D}^*$ pole, its formation dynamics inevitably involves a significant coupling with the quark-model state. }

{ Additionally, we have examined a wide range of bare mass $m_0$ values and its influence on the pole behaviors. The observed behavior is quite systematic: when the bare mass is closer to the open-charm threshold, the system tends to favor a strong-coupling mode. For example, when $m_0$ is set very close to 3872 MeV, the bare state is easily pulled toward the negative imaginary axis of the momentum $k$-plane and becomes a virtual state. In this case, its shadow pole associated with the anti-resonance of $\chi_{c1}^{\prime}$ moves to the physical sheet and forms a bound state corresponding to $X(3872)$. For the $\chi_{c1}^{\prime}$ pole as a virtual state, a cusp-like structure may emerge near the threshold, although its signal becomes weaker as the pole moves farther from threshold. Meanwhile, the $D\bar{D}^*$ pole in the complex plane appears far from the physical region, making it effectively unobservable in experiments. This prediction is clearly inconsistent with the experimentally observed $\chi_{c1}(4010)$ state.}

\section{The uniformization in the two-channel problem}

For the two-channel threshold $m_{th0}$ and $m_{th\pm}$, we define two momentum-like  variables 
\begin{eqnarray}
p_0^2=E^2-m_{th0}^2, ~~~p_{\pm}^2=E^2-m_{th\pm}^2, 
\end{eqnarray}
and $p_{\pm}^2-p_0^2=\Delta^2$. The operation mapping $E$ to $z$ is given by the following relations,
\begin{eqnarray}
p_{\pm}+p_{0}=\Delta z, ~~~ p_{0}-p_{\pm}=\frac{\Delta}{z}.
\end{eqnarray}
Based on the definition of the variable $z$, the same energy $E$ in four different Riemann sheets can be mapped onto different regions in the $z$-plane, which is the so-called uniformization process. Here, the origin of the $z$-plane is mapped to the infinity of the $(-,+)$ and $(+,-)$ Riemann sheets, so the molecular pole and the $\chi_{c1}^{\prime}$ state stem from the origin and the real axis, respectively.

\section{The non-open charm and open charm decay behaviors of bare $P$-wave charmonium $\chi_{c1}^{\prime}$}

The main non-open-charm modes include the radiative and {light hadron decays} and their rates are usually related to the radial wave function of $\chi_{c1}^{\prime}$, which can be described by a simple harmonic oscillator function \cite{Close:2005se,Guo:2022zbc}, i.e., 
\begin{align}
R_{nL}(r,\beta)=\beta^{\frac{3}{2}}\sqrt{\frac{2n!}{\Gamma(n+L+\frac{3}{2})}}(\beta r)^Le^{\frac{-r^2\beta^2}{2}}L_n^{L+\frac{1}{2}}(r^2\beta^2), 
\end{align}
where $L_n^{L+\frac{1}{2}}$ is the associated Laguerre polynomial with the orbital angular momentum $L$ and the radial quantum number $n$.  Under the heavy quark symmetry, it can be concluded that the $P$-wave charmonium triplets share the same radial wave function, so the unique unknown parameter $\beta=0.7\sim0.9$ GeV \cite{Guo:2022zbc} can be reliably determined by the experimental widths of $\chi_{c2}^{\prime}$ and $\chi_{c0}^{\prime}$ \cite{ParticleDataGroup:2022pth}. Subsequently, the decays into light hadrons or a charmonium with a photon can be calculated.

The important non-open charm decays of the $c\bar{c}$ state include the light hadron decay and the radiative transition into the lower charmoniums. For a $n^3P_1$ charmonium, its inclusive decays to various light hadron final states mainly occur through the annihilation process $n^3P_1 \to q\bar{q}g$, whose width depends on the first-order derivative of its radial wave function at the origin \cite{Kwong:1987ak}, 
\begin{eqnarray}
\Gamma(n^3P_1 \to q\bar{q}g)=\frac{32\alpha_s^3}{9\pi m_c^4}|R^{\prime}_{nP}(0)|^2 \text{ln}(m_c \left < R \right >),
\end{eqnarray}
where $\left < R \right > $ is the average radius of the $n^3P_1$ state, $m_c=1.65$ GeV and $\alpha_s=0.26$.  For the radiative decays of the charmonium, the partial width of the $E1$ transition $n^{2S+1}L_J \to n^{\prime2S+1}L^{\prime}_{J^{\prime}} \gamma$ is given by \cite{Kwong:1988ae},
\begin{eqnarray}
\Gamma_{E1}=\frac{4}{3}\alpha e_c^2 \omega^3\delta_{L,L'\pm1}C_{if}\left|\langle \psi_f|r|\psi_i\rangle\right|^2,
\end{eqnarray}
with
\begin{eqnarray}
C_{if}=max(L,L')(2J'+1) {\begin{Bmatrix} L'&J'&S\\ J&L&1 \end{Bmatrix}}^2,
\end{eqnarray}
where the $e_c$ is the charm quark charge in units of $|e|$, $\alpha$ is the fine-structure constant, $\omega$ is the emitted photon energy and $\langle \psi_f|r|\psi_i\rangle=\int_0^{\infty}R_{n'L'}(r)rR_{nL}(r)r^2dr$ is the transition matrix element. 
The partial width of the $M1$ radiative transition $n^{2S+1}L_J \to {n'}^{2S'+1}{L}_{J'} \gamma$ with the spin flip can be written as \cite{Novikov:1977dq}
\begin{equation}
\Gamma_{M1}=\frac{4\alpha e_c^2\omega^3}{3m_c^2}\delta_{S,S'\pm1}\frac{2J'+1}{2L+1}\left|\left< \psi_f\left|j_0\left(\frac{\omega r}{2}\right)\right|\psi_i\right>\right|^2,
\end{equation}
with
\begin{equation}
\left< \psi_f\left|j_0\left(\frac{\omega r}{2}\right)\right|\psi_i\right>=\int_0^{\infty}R_{n'L'}(r)j_0\left(\frac{\omega r}{2}\right)R_{nL}(r)r^2dr,
\end{equation}
where  $j_0(\frac{\omega r}{2})$ is the spherical Bessel function. For the first excited $P$-wave charmonium $\chi_{c1}^{\prime}$, the kinematically allowed final states of $E1$ transition include $J/\psi \gamma$, $\psi(3686)\gamma$, $\psi(3770)\gamma$ and $\psi_2(3823) \gamma$, and the corresponding $M1$ transition process is $\chi_{c1}^{\prime} \to h_c \gamma$.

The quark pair creation (QPC) model is a very successful phenomenological approach in depicting the two-body OZI-allowed strong decay behaviors covering the light meson to heavy quarkonium system \cite{Micu:1968mk,LeYaouanc:1972vsx}. The QPC model assumes that the quark-antiquark pair $q\bar{q}$ created from the vacuum is a $ ^3P_0$ state with the spin-parity $J^{PC}=0^{++}$, and the transition operators $\mathcal{T}$ can be expressed as
\begin{eqnarray}
\mathcal{T}& = &-3\gamma \sum_{m}\langle 1m;1-m|00\rangle\int d \textbf{p}_3d\textbf{p}_4 \delta ^3 (\textbf{p}_3+\textbf{p}_4) \nonumber \\
&& \times \mathcal{Y}_{1m}\left( \frac{\textbf{p}_3-\textbf{p}_4}{2} \right) \chi _{1,-m}^{34} \phi _{0}^{34}
\omega_{0}^{34} b_{3i}^{\dag} (\textbf{p}_3) d_{4j}^{\dag}(\textbf{p}_4), \nonumber \\
\end{eqnarray}
where $\mathcal{Y}_{lm}\left( \textbf{p}\right)=p^lY_{lm}(\theta_p,\phi_p)$, $b_3^{\dag}(d_{4}^{\dag})$ is the quark (antiquark) creation operator, and $\phi _{0}^{34}=(u\bar{u}+d\bar{d}+s\bar{s})/\sqrt{3}$ and $\omega_{0}^{34} $ are SU(3) flavor and color wave function of vacuum quark pair, respectively. The dimensionless parameter $\gamma^2$, reflecting the strength of the vacuum quark pair creation, has been determined as $40.9\pm8.2$ in Ref. \cite{Guo:2022zbc} for the charmonium family. The transition amplitude of $\chi_{c1}^{\prime} \to D\bar{D}^*$ is
\begin{eqnarray}
\mathcal{M}=\left< D\bar{D}^* |\mathcal{T} | \chi_{c1}^{\prime} \right>,
\end{eqnarray}
which is proportional to the overlap integral of the wave functions in momentum space
\begin{eqnarray}
I(\textbf{q})&=&\int d^3\textbf{p} \Psi_{n_AL_AM_{L_A}}(\textbf{q}+\textbf{p}) \Psi^\ast _{n_BL_BM_{L_B}}(\frac{m_q \textbf{q}}{m_c+m_q}+\textbf{p})  \nonumber \\
&&  \times \mathcal{Y}_{lm}(\textbf{p})
\Psi^\ast _{n_CL_CM_{L_C}}(\frac{m_q \textbf{q}}{m_{\bar{c}}+m_q}+\textbf{p}),   
\end{eqnarray}
where $\textbf{q}$ denotes the momentum of either outgoing charmed meson, and $\Psi_{nLM_L}(\textbf{p})$ is the spatial wave function, while the notation $A$, $B$ and $C$ refer to the $\chi_{c1}^{\prime}$, $D$ and $\bar{D}^*$ states, respectively. It is worth emphasizing that the overlap integral $I(\textbf{q})$ provides the momentum-dependent part of the $\mathcal{V}_{0j}(\textbf{q})$ and $\mathcal{V}_{i0}(\textbf{q}^{\prime})$ defined in the main text.

The numerical coupled-channel interaction $\mathcal{V}_{0j}(\textbf{q})$ between the bare $\chi_{c1}^{\prime}$ and the $D^0\bar{D}^{*0}$/$D^+\bar{D}^{*-}$ continuum  estimated by the QPC model is shown in Fig.~\ref{Appen:fig1} with $\beta=$0.7, 0.8 and 0.9 GeV. $\beta$ is a quantity reflecting the wave function distribution of $\chi_{c1}^{\prime}$.  We also illustrate the variation of the total decay width of $\chi_{c1}^{\prime} \to D\bar{D}^*+c.c.$ with $\beta$, which is estimated to be $250\sim320$ MeV. Such a large decay width also implies that the $s$-channel coupling with the $D\bar{D}^*$ may be too strong to be ignored.  

Based on the same range of  $\beta$, the estimated widths of the radiative transitions and inclusive light hadron decays of the bare $\chi_{c1}^{\prime}$ are presented in  Fig.~\ref{Appen:fig2}. It can be seen that the $\psi(3686)\gamma$ is the dominant radiative decay mode although its phase space is far smaller than that of $J/\psi \gamma$, which mainly benefits from the node effect from the wave function overlap. Additionally, the $\psi(3770)\gamma$ and $\psi_2(3823) \gamma$ are also important radiative modes, whose widths are $10\sim20$ keV.  The $\Gamma(\chi_{c1}^{\prime} \to q\bar{q}g)$ can reach $2.6\sim7.6$ MeV, which should govern the non-open-charm decays of the bare $\chi_{c1}^{\prime}$.
If assuming the $D\bar{D}^*$ component of $X(3872)$ does not decay to $\psi(3686)\gamma$ and $J/\psi \gamma$, the theoretical value of the ratio $R=\Gamma(\psi(3686)\gamma)/\Gamma(J/\psi \gamma)$ including the phase space correction coincides with the corresponding experimental values from the Belle and LHCb Collaboration \cite{Belle:2011wdj,LHCb:2014jvf} within a narrower range of $\beta=0.73\sim0.766$ GeV. When adopting $\beta=0.75$ GeV, two typical width values of $\Gamma_a=3488$ keV and $\Gamma_b=220$ keV, which correspond to the light hadron decays and radiative decays of the bare $\chi_{c1}^{\prime}$, can be obtained.

\begin{figure}[t]
\centerline{\includegraphics[width=7.0cm]{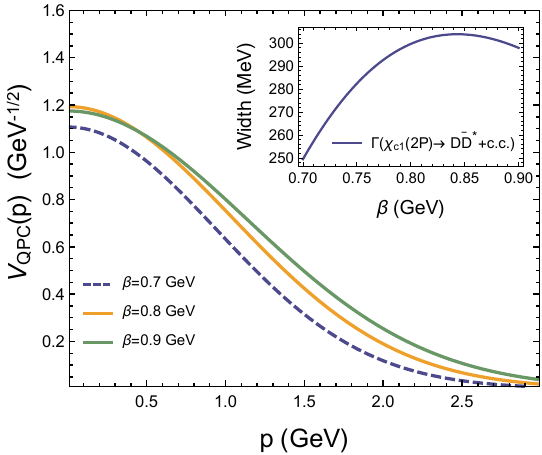}}
	\caption{ The numerical $\mathcal{V}_{0j}(\textbf{q})$ between the bare $\chi_{c1}^{\prime}$ and the $D^0\bar{D}^{*0}$/$D^+\bar{D}^{*-}$ continuum and the corresponding open-charm decay width in the QPC model.  \label{Appen:fig1} }
\end{figure}

\begin{figure*}[t]
\centerline{\includegraphics[width=18.0cm]{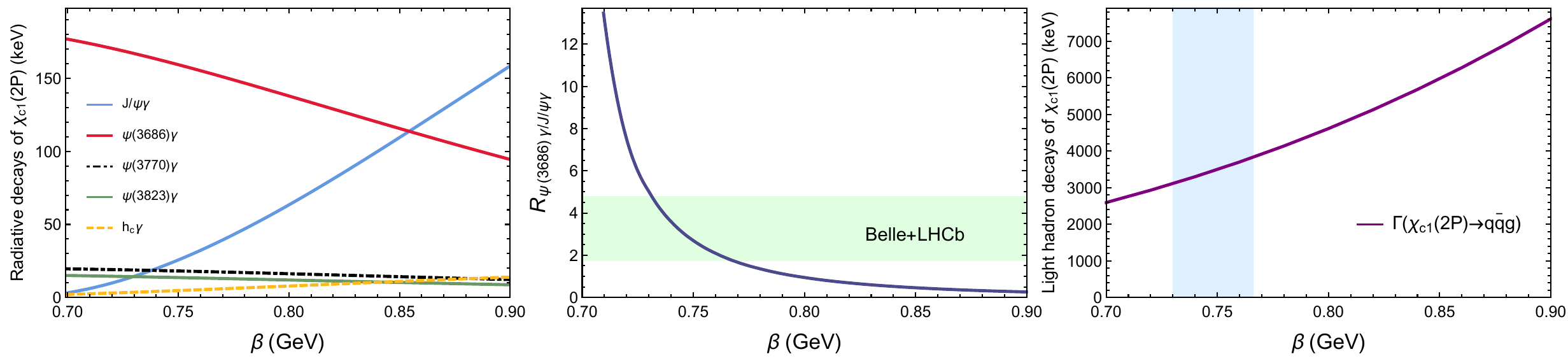}}
	\caption{ Theoretical widths of the radiative transitions and inclusive light hadrons decays of the bare $\chi_{c1}^{\prime}$ state.  \label{Appen:fig2} }
\end{figure*}

\section{The three-body $D\bar{D}\pi$ threshold dynamics and the self-energy effect of $\bar{D}^*$ in the $D\bar{D}^*$ scattering}

A typical OPE potential $V_{\pi}$ of the $[D\bar{D}^*]_i \to [D\bar{D}^*]_j$ can be written as
\begin{eqnarray}
 V_{\pi}^{ij}(q,q^{\prime},z)=\frac{g^2}{4f_{\pi}^2}\frac{(\varepsilon \cdot p)(\varepsilon^{\prime} \cdot p)}{p_0^2-(q^2+q^{\prime 2}-2qq^{\prime} z)-m_{\pi}^2+i\epsilon},  \nonumber \\
\end{eqnarray}
where $i(j)=1, 2$ stands for $[D^0\bar{D}^{*0}]$ and $[D^+\bar{D}^{*-}]$, respectively, and the $p$ is defined as the four-momentum ($p_1$-$p_4$) of the exchanged pion, and $p_1$, $p_2$, $p_3$ and $p_4$ correspond to the four-momentum of the initial charmed meson $D$, $\bar{D}^*$ and final charmed meson $\bar{D}$ and $D^*$, respectively. The $q(q^{\prime})$ and $z$ denote the magnitudes of the three-momentum $\boldsymbol{p}_1$($\boldsymbol{p}_4$) and the cosine of the angle between $\boldsymbol{p}_1$ and $\boldsymbol{p}_4$, respectively. The coupling constant $g=0.53$ \cite{ParticleDataGroup:2022pth} and pion decay constant $f_{\pi}=0.086$ \cite{Lin:2022wmj}.  Since the $p_0\sim(m_{D^*}-m_D)>m_{\pi}^{\mathrm{phy}}$ will lead to an on-shell pion exchange, the $D\bar{D}^*$ system couples to the three-body channel of $D\bar{D}\pi$. In order to include this three-body threshold effect, the OPE potential can be modified as \cite{Lin:2022wmj}
\begin{eqnarray}
&&V_{\pi}^{ij}(q,q^{\prime},z)= \nonumber \\
&&\frac{g^2}{4f_{\pi}^2} \frac{(\varepsilon \cdot p)(\varepsilon^{\prime} \cdot p)}{(E^{\prime}+\delta_{ij})^2-(q^2+q^{\prime 2}-2qq^{\prime} z)-m_{\pi}^2+i\epsilon},  \label{Appen:eq10}
\end{eqnarray}
where the system energy $E$ is the sum of the center-of-mass kinetic energy $E^{\prime}=k_0^2/(2\mu)$ and the lowest two-body threshold $m_{D^0}+m_{\bar{D}^{*0}}$, and $\delta_{11}=m_{D^{*0}}-m_{D^0}$, $\delta_{12}=m_{D^{*0}}-m_{D^+}$, $\delta_{21}=m_{D^{*0}}-m_{D^+}$ and $\delta_{22}=m_{D^{*0}}+m_{D^0}-2m_{D^+}$.  {  The pole of $V_{\pi}^{ij}(q,q^{\prime},z)$ corresponds to the on-shell condition of $D\bar{D}\pi$ three-body intermediate state, where the $D$ and $\bar{D}$ are on shell because they are in the initial or final state.
After the integral of the partial-wave projection, this pole results in the discontinuity from the $D\bar{D}\pi$ threshold, i.e. the three-body unitary cut. It can be seen from Eq. (\ref{Appen:eq10}) that when $p,p^{\prime}
\to 0$,
\begin{eqnarray}
    V^{11}_{\mathrm{OPE}}(p,p^{\prime})&\propto&((k_0^2/(2\mu)+\delta+m_{\pi})((k_0^2/(2\mu)+\delta-m_{\pi}))^{-1} \\
    &\propto&((E-m_D^0-m_D^0-m_{\pi^0})(E-m_D^0-m_D^0+m_{\pi^0}))^{-1}, \nonumber
\end{eqnarray}
 where a singularity at $E=m_D+m_D+m_{\pi}$ appears. 
 
 The $S$-matrix unitarity requires that the imaginary part of the revised OPE potential corresponds to the phase space of the on-shell $D\bar{D}\pi$ three-body state, which can be seen from
\begin{eqnarray}
    \text{Im }V_{\pi}^{ij}(q,q^{\prime},z)\propto
    &&\,\delta\left((E^{\prime}+\delta_{ij})^2-(q^2+q^{\prime 2}-2qq^{\prime} z)-m_{\pi}^2+i\epsilon\right). 
\end{eqnarray}
 Therefore, our revised partial-wave OPE potential by performing the integral for the invariant phase space naturally induces a three-body threshold cut. Additionally, the complete three-body unitarity of $D\bar{D}\pi$ implies the inclusion of the self-energy effect of $\bar{D}^*$ as explained in Ref. \cite{Mai:2017vot}, and its contribution is also considered in our calculation. We also further carry out a scattering resummation of the revised OPE potential via a dynamical equation such as the Lippmann-Schwinger equation (LSE) or the Schrödinger equation involving the self-energy effect of $\bar{D}^*$, which ensures the three-body unitarity in our formalism.} 
 
Here, for the concerned poles near the two-body threshold, ignoring the kinetic energy of the heavy meson is a good approximation. When considering the kinetic energy term of heavy meson, Eq.~(\ref{Appen:eq10}) can be further revised by the following replacement of $\delta \to \delta-(p^2+p^{\prime2})/(2m_{D})$.  More discussions and applications of the three-body threshold effects can be found in Refs. \cite{Lin:2022wmj,Wang:2023iaz}.

The matrix form of the contact interaction of the $D\bar{D}^*$ scattering reads
\begin{eqnarray}
V_{\text{contact}}=\begin{pmatrix}
   C_t & C_t^{\prime}   \\
    C_t^{\prime}  & C_t   \\
\end{pmatrix}.
\end{eqnarray}
We define two new leading-order low-energy coupling constant $V_{ct}^{I=0}$ and $V_{ct}^{I=1}$, which can be related to $C_t$ and $C_t^{\prime}$ by 
\begin{eqnarray}
C_t&=&\frac{1}{2}(V_{ct}^{I=0}+V_{ct}^{I=1}), \nonumber \\
C_t^{\prime}&=&\frac{1}{2}(V_{ct}^{I=0}-V_{ct}^{I=1}).
\end{eqnarray}
Here, the $I$ represents the isospin of the $D\bar{D}^*$ system.

Because the $D^{*0}$ and $D^{*\pm}$  have comparable widths with the pole width of $X(3872)$, it is necessary to take into account the width for the propagator of $D^*$ \cite{Du:2021zzh}, which includes both the intermediate $\bar{D}\pi$ and $\bar{D}\gamma$ self-energy diagrams and  is different from the mechanism of the revised OPE potential with the three-body threshold cut. The real part of the self-energy of $D^*$ is approximately absorbed into the physical mass of the $D^*$ meson. The imaginary part of the self-energy of $D^*$ is given by
\begin{eqnarray}
\Gamma_{D^{*+}}(E)&=&\frac{g^2m_{D^0}}{12\pi f_{\pi}^2m_{D^{*+}}}k^3_{D^0\pi^+}+\frac{g^2m_{D^+}}{24\pi f_{\pi}^2m_{D^{*+}}}k^3_{D^+\pi^0}\nonumber \\
&&+\Gamma(D^{*+} \to D^{+} \gamma), \nonumber \\
\Gamma_{D^{*0}}(E)&=&\frac{g^2m_{D^0}}{24\pi f_{\pi}^2m_{D^{*0}}}k^3_{D^0\pi^0},
\end{eqnarray}
where the $D^*$ width is treated as a function of the center-of-mass energy $E$ rather than a constant since the $D^*$ is not always on-shell in the complex energy space. Consequently, the the width of $D^*$ can be introduced into the scattering dynamics by modifying the Schrödinger equation as \cite{Lin:2022wmj}
\begin{eqnarray}
    E\phi(\bm{p})=(\frac{\bm{p}^2}{2\mu}-i\frac{\Gamma(E)}{2})\phi(\bm{p})+\int \frac{d^3\bm{k}}{(2\pi)^3} V(\bm{p},\bm{{k}})\phi(\bm{k}), \nonumber \\
\end{eqnarray}
where $\frac{\bm{p}^2}{2\mu}-i\frac{\Gamma(E)}{2}$ means a modified kinetic energy term for the unstable system.

\section{The pole behaviors of the full $D\bar{D}^*$ scattering dynamics by considering the isospin violation of the contact potential }

Considering the full $D\bar{D}^*$ scattering dynamics and the evident isospin violation of the contact interaction, the pole behaviors of the $X(3872)$ and the new resonance are summarized in Table \ref{tab:2} and Table \ref{tab:3}, which correspond to $V_{ct}^{I=1}=30$ $\text{GeV}^{-2}$ and $V_{ct}^{I=1}=-30$ $\text{GeV}^{-2}$, respectively. Interestingly, there almost exists a one-to-one correspondence between the cutoff $\lambda$ and the pole imaginary of the new charmoniumlike resonance, which almost does not depend on the contact potential $V_{ct}^{I=1}$.

Furthermore, it can be seen that the isospin violation impact on $X(3872)$ is obviously greater than that on the dressed $\chi_{c1}^{\prime}$ resonance or the distorted $D\bar{D}^*$ resonance. This feature can explain the absence of the experimental signal of this higher charmoniumlike resonance in the final states of $J/\psi \pi^+\pi^-$ \cite{ParticleDataGroup:2022pth}, which is a typical isospin violation channel of discovering the $X(3872)$. 

\begin{table*}[t]
\renewcommand{\arraystretch}{2}
    \centering
\caption{ The poles and component possibilities of $X(3872)$ and a new resonance in the complete coupled-channel dynamics. The pole of $X(3872)$ is given by the binding energy relative to the $D^0\bar{D}^{*0}$ threshold. The $\chi$ and $M$ superscripts denote the dressed $\chi_{c1}^{\prime}$ resonance and the distorted  $D\bar{D}^*$ resonance in the $(-,-)$ sheet, respectively. Here, $V_{ct}^{I=1}=30$ $\text{GeV}^{-2}$ and all pole positions are in units of MeV.  }
\setlength{\tabcolsep}{1.6mm}{
    \begin{tabular}{c|c|c|ccccccc}
    \hline
     \hline
     \multicolumn{3}{c|}{$\lambda$ ($V_{ct}^{I=0}$ ) ~~$\text{GeV}^{-1}$ $(\text{GeV}^{-2})$ }  & 0.5 (96.4)  & 1.0 (21.8) & 1.25 (10.3)  &1.5 (2.7) & 2.5 (-14.4)  \\
    \hline
    \multirow{8}{*}{With $\Gamma_a+\Gamma_b$}     & \multirow{4}{*}{$X(3872)$} &   Pole & -0.077-1.37i  &  -0.064-0.333i &  -0.066-0.189i &  -0.074-0.133i  &  -0.075-0.069i      \\
    \cline{3-8}
    &  &  $\mathcal{P}_{D^0\bar{D}^{*0}}$ & 0.138-0.010i  &  0.681-0.148i &  0.809-0.095i &  0.858-0.061i  &  0.916-0.018i    \\
    \cline{3-8}
     &  &  $\mathcal{P}_{D^+\bar{D}^{*-}}$ & 0.116+0.000i  &  0.130+0.057i &  0.098+0.047i &  0.083+0.034i  &  0.063+0.013i    \\
    \cline{3-8}
     &  &  $\mathcal{P}_{\chi_{c1}^{\prime}}$ & 0.746+0.010i  &   0.189+0.091i &  0.093+0.048i &  0.059+0.027i  &  0.021+0.005i    \\
    \cline{2-8}
      & \multirow{4}{*}{New resonance} & Pole &  4149-141i$^{M}$  &  4061-129i$^{M}$  &  4023-92i$^{\chi}$ &  4004-56i$^{\chi}$ &  3977-8i$^{\chi}$  \\
      \cline{3-8}
    &  &  $\mathcal{P}_{D^0\bar{D}^{*0}}$ & 0.324+0.054i   &  0.263-0.181i &  0.205-0.230i &  0.133-0.198i  &  0.082-0.037i    \\
    \cline{3-8}
    &  &  $\mathcal{P}_{D^+\bar{D}^{*-}}$ & 0.328+0.001i   &  0.273-0.236i &  0.154-0.335i &   0.070-0.226i  &  0.073-0.055i    \\
    \cline{3-8}
     &  &  $\mathcal{P}_{\chi_{c1}^{\prime}}$ & 0.348-0.055i   &  0.464+0.417i &  0.641+0.565i &  0.797+0.424i &  0.845+0.092i    \\
    \hline
     \hline
    \end{tabular}
    }
    \label{tab:2}
\end{table*}

\begin{table*}[t]
\renewcommand{\arraystretch}{2}
    \centering
\caption{ The poles and component possibilities of $X(3872)$ and a new resonance in the complete coupled-channel dynamics. The pole of $X(3872)$ is given by the binding energy relative to the $D^0\bar{D}^{*0}$ threshold. The $\chi$ and $M$ superscripts denote the dressed $\chi_{c1}^{\prime}$ resonance and the distorted  $D\bar{D}^*$ resonance in the $(-,-)$ sheet, respectively. Here, $V_{ct}^{I=1}=-30$ $\text{GeV}^{-2}$ and all pole positions are in units of MeV.  }
\setlength{\tabcolsep}{1.6mm}{
    \begin{tabular}{c|c|c|ccccccc}
    \hline
     \hline
     \multicolumn{3}{c|}{$\lambda$ ($V_{ct}^{I=0}$ ) ~~$\text{GeV}^{-1}$ $(\text{GeV}^{-2})$ }  & 0.5 (97.0)  & 1.0 (25.4) & 1.25 (14.6)  &1.5 (7.7) & 2.5 (-8.0)    \\
    \hline
    \multirow{8}{*}{With $\Gamma_a+\Gamma_b$}     & \multirow{4}{*}{$X(3872)$} &   Pole & -0.064-1.32i  &  -0.067-0.178i &  -0.078-0.113i &  -0.073-0.084i  &  -0.076-0.055i    \\
    \cline{3-8}
    &  &  $\mathcal{P}_{D^0\bar{D}^{*0}}$ & 0.170-0.041i  &  0.905-0.067i &  0.946-0.027i &  0.964-0.013i  &  0.982-0.003i    \\
    \cline{3-8}
     &  &  $\mathcal{P}_{D^+\bar{D}^{*-}}$ & 0.092+0.009i  &  0.015+0.014i &  0.010+0.006i &  0.008+0.003i  &  0.006+0.001i    \\
    \cline{3-8}
     &  &  $\mathcal{P}_{\chi_{c1}^{\prime}}$ & 0.738+0.032i  &   0.085+0.053i &  0.044+0.021i &  0.028+0.010i  &  0.012+0.002i    \\
    \cline{2-8}
      & \multirow{4}{*}{New resonance} & Pole &  4150-141i$^{M}$  &  4066-127i$^{M}$  &  4031-94i$^{\chi}$ &  4008-60i$^{\chi}$ &  3978-9i$^{\chi}$   \\
      \cline{3-8}
    &  &  $\mathcal{P}_{D^0\bar{D}^{*0}}$ & 0.329+0.043i   &  0.284-0.182i &  0.230-0.258i &  0.135-0.228i  &  0.075-0.056i    \\
    \cline{3-8}
    &  &  $\mathcal{P}_{D^+\bar{D}^{*-}}$ & 0.323+0.011i   &  0.269-0.197i &  0.217-0.271i &   0.129-0.243i  &  0.081-0.073i    \\
    \cline{3-8}
     &  &  $\mathcal{P}_{\chi_{c1}^{\prime}}$ & 0.348-0.054i   &  0.447+0.379i &  0.553+0.529i &  0.736+0.471i &  0.844+0.129i    \\
    \hline
     \hline
    \end{tabular}
    }
    \label{tab:3}
\end{table*}

\end{onecolumngrid}

\end{document}